\documentclass[reprint,amsmath,amssymb,prxlife,floatfix]{revtex4-2}
\usepackage{graphicx}     
\usepackage{dcolumn}      
\usepackage{bm}           
\usepackage{hyperref}     
\usepackage{orcidlink}    
\usepackage[mathlines]{lineno}
\usepackage{mathtools}
\usepackage{subcaption}
\usepackage[percent]{overpic} 

\usepackage{float}

\begin{document}


\title{Stochasticity and Environmental Switching Shape Quorum Sensing Evolution in Bacterial Populations}

\author{Uttam Kumar$^{1}$\,\orcidlink{0000-0002-0751-9388}} \email{uttam@as.edu.tw, uttamup11@gmail.com}
\author{Hong-Yan Shih$^{1,2}$\,\orcidlink{0000-0002-3578-7785}} \email{hongyan@as.edu.tw}
\affiliation{%
 $^1$Institute of Physics, Academia Sinica, Taipei 115201, Taiwan\\
 $^2$Physics Division, National Center for Theoretical Sciences, Taipei 106319, Taiwan 
}%

\begin{abstract}
Quorum sensing (QS) is classically understood as a mechanism by which bacterial cells sense their own density, but environmental fluctuations make
any single cell's estimate unreliable. An alternative view holds that populations pool individual estimates to reach reliable collective decisions.
We investigate how a two-autoinducer toggle switch network enables this collective adaptation in fluctuating environments. At the single-cell level,
the toggle switch is bistable. Membrane permeability controls autoinducer secretion and is the evolving trait. Using an individual-based model of a spatially structured population, we show that permeability evolves through a collective mechanism. Environmental frequency is the primary determinant of evolutionary outcome.
Under strong asymmetry, the dominant trait reaches a high plateau. The trait linked to the rare environment stays low.
It rises smoothly as its environment becomes more frequent, with no sharp threshold. Total permeability stays close to a fixed budget; environmental frequency mainly sets how it is split between the two traits.
When the two environments are equally frequent, no single trait wins. Both traits coexist, and noise asymmetry between the two sensing channels biases
the levels at which they coexist rather than selecting a winner. Coexistence is also the slowest outcome to reach.
Sensing noise governs a second axis of the outcome. Higher noise speeds adaptation, but the evolved permeability is non-monotonic in noise, peaking at
an intermediate value. Earlier work found that more noise favors collective sensing as long as cells are correct on average.
We develop a two-trait adaptive-dynamics model that unifies these regimes. Away from symmetry, it reduces to a closed one-dimensional equation that captures the individual-based results. At symmetry, weak cross-repression gives a single coexistence attractor, consistent with replicate populations maintaining both traits.
\end{abstract}

\maketitle

\section{\label{sec:level1}Introduction}
Bacteria live in rapidly fluctuating environments in nature \cite{Cao2017Ready, Kussell2005}. They constantly adjust their physiology to survive, for example, by forming biofilms \cite{Flemming2016, HallStoodley2004}, tolerating antibiotics \cite{Balaban2004}, or hedging between alternative phenotypes \cite{Beaumont2009}. Each switch is a decision about the current
environment, and a single cell makes it from a noisy estimate. One of the key strategies they use is quorum sensing (QS), in which cells release and detect signaling molecules called autoinducers (AIs) to coordinate collective behaviors \cite{Miller2001, Waters2005, Papenfort2016}. When AI concentration crosses a threshold, cells alter their gene expression in a coordinated way~\cite{Miller2001, Papenfort2016}.
Both Gram-negative and Gram-positive bacteria use QS, through acyl-homoserine lactones and oligopeptides respectively \cite{Waters2005, Ng2009}. It regulates biofilm formation, virulence, competence, and sporulation \cite{Rutherford2012, Whiteley2017, Mukherjee2019}. 


Many of these programs are mutually exclusive, so cells must choose between them. \textit{Bacillus subtilis} uses two peptide signals, ComX and CSF, to choose between competence and sporulation under stress \cite{Schultz2009, Bareia2017}. \textit{Vibrio cholerae} expresses virulence and biofilm genes at low cell density and represses them at high density \cite{Jemielita2018}. Such choices often appear as bimodal gene expression. In \textit{B. subtilis}, positive feedback produces bimodal competence and sporulation \cite{Maamar2005, Veening2005}. 

The simplest circuit with two exclusive states is the genetic toggle switch, a pair of mutually repressing genes, first built in \textit{E. coli} \cite{Gardner2000}. Kobayashi \textit{et al.} coupled a toggle to a QS input, giving cells a heritable binary response to a transient signal \cite{Kobayashi2004}. QS coupling can also synchronize gene circuits across a population \cite{Danino2010}. Models show that it can align toggle states \cite{Nikolaev2016, Weber2016} or instead produce bimodal and multimodal responses \cite{Quan2016, Wu2014quorum}. How the coupling evolves when cells must choose between two programs in a fluctuating environment remains unaddressed.

The classical interpretation of QS proposes that bacteria monitor AI concentrations to measure population density, thus activating specific functions only when a critical threshold is reached \cite{Waters2005}. This fits public goods, such as extracellular matrix, which are costly for one cell to produce but effective when many cells produce them \cite{Darch2012, Mellbye2014}. It fits poorly when QS regulates private traits, such as antibiotic tolerance or metabolic adjustment, which benefit only the producing cell \cite{Schuster2017, Leung2012, An2014}.  Since these traits do not depend on neighboring cells, it is unclear why bacteria would need to sense population density to control them. 
Signal accumulation also depends on more than density. Diffusion rates and pH change autoinducer levels independently of cell number \cite{Decho2010, West2012, Cornforth2014}. Redfield therefore proposed that cells sense diffusion rather than density \cite{Redfield2002}, and Hense \textit{et al.} combined density, diffusion and spatial distribution into a single measure of signaling efficiency \cite{Hense2007}. 

There is growing support for the view that individual cells make imperfect estimates of the environment. By sharing AIs, they pool these estimates and reach more accurate collective decisions~\cite{Keller2006, MorenoGamez2023}.
Pooling of noisy individual information also underlies collective decisions in animal groups and social networks \cite{Berdahl2013, Golub2010}. Moreno-G\'amez \textit{et al.} showed that such pooling can drive the evolution of QS \cite{MorenoGamez2023}. In their
model, cells evolve the membrane permeability of a single autoinducer in two equally frequent environments. More noise in individual estimates favors collective sensing, provided cells are correct on average. Salahshour and colleagues studied noise in communication rather than sensing. Production noise degrades collective inference \cite{Salahshour2019SciRep}, and an intermediate communication noise maximizes responsiveness \cite{Salahshour2019PRL}. In their models, noise levels are fixed rather than evolved. Here, we ask how sensing noise, environmental asymmetry, and competition between two signals jointly shape what evolves.

We study a two-autoinducer toggle switch in which each autoinducer is linked to a distinct phenotypic program. Its exchange with the environment is set by a membrane permeability that evolves independently for each signal. The two signals compete within each cell. This competition raises questions that a single-signal model cannot pose: how unequal environmental frequencies bias the outcome, and whether one signal wins or both persist. We embed the switch in a spatially structured, evolving population. Environmental frequency sets how a nearly fixed total permeability is divided between the two traits. The
trait tied to a rare environment is suppressed but not lost, and it rises smoothly as its environment becomes more frequent. Sensing noise has a non-monotonic effect. An intermediate noise level maximizes the evolved permeability, and noise controls adaptation speed and plateau height separately. When the two environments are equally frequent, no single trait wins. Both traits coexist, and noise asymmetry grades the coexistence level.
An adaptive-dynamics model reproduces these outcomes and links single-cell bistability to population-level evolution.

The paper is organized as follows.
Section~\ref{sec:model} describes the toggle-switch equations and the individual-based model. Section~\ref{sec:results} presents the simulation results under varying environmental conditions. Section~\ref{sec:theory} develops an adaptive-dynamics model of phenotypic evolution. Section~\ref{sec:discussion} discusses the biological implications of these findings and limitations. Finally, Section~\ref{sec:conclusions} summarizes the main findings.
\section{Model Description}
\label{sec:model}
We consider a system in which bacteria produce two autoinducers, $a_{1}$ and $a_{2}$, corresponding to two distinct characteristics under different environmental conditions. This is analogous to natural systems such as \textit{Bacillus subtilis}, which produces two autoinducing peptides that regulate two developmental pathways: competence and sporulation. We consider a population of $N = 2500$ bacteria arranged on a $50 \times 50$ grid with periodic (toroidal) boundary conditions, so that every cell has an equivalent neighborhood and boundary effects do not bias the collective dynamics. Each cell carries two heritable membrane permeability traits, $c_1$ and $c_2$, initialized to zero at generation $t = 0$. The environment alternates randomly each generation between two states, $E_1$ (probability $p_1$) and $E_2$ (probability $p_2 = 1 - p_1$), which impose distinct selection pressures on autoinducer production.
\subsection{Single-cell toggle switch  and bistability}
\label{sec:toggle}
\begin{figure}[ht]
\centering
\includegraphics[width=\linewidth]{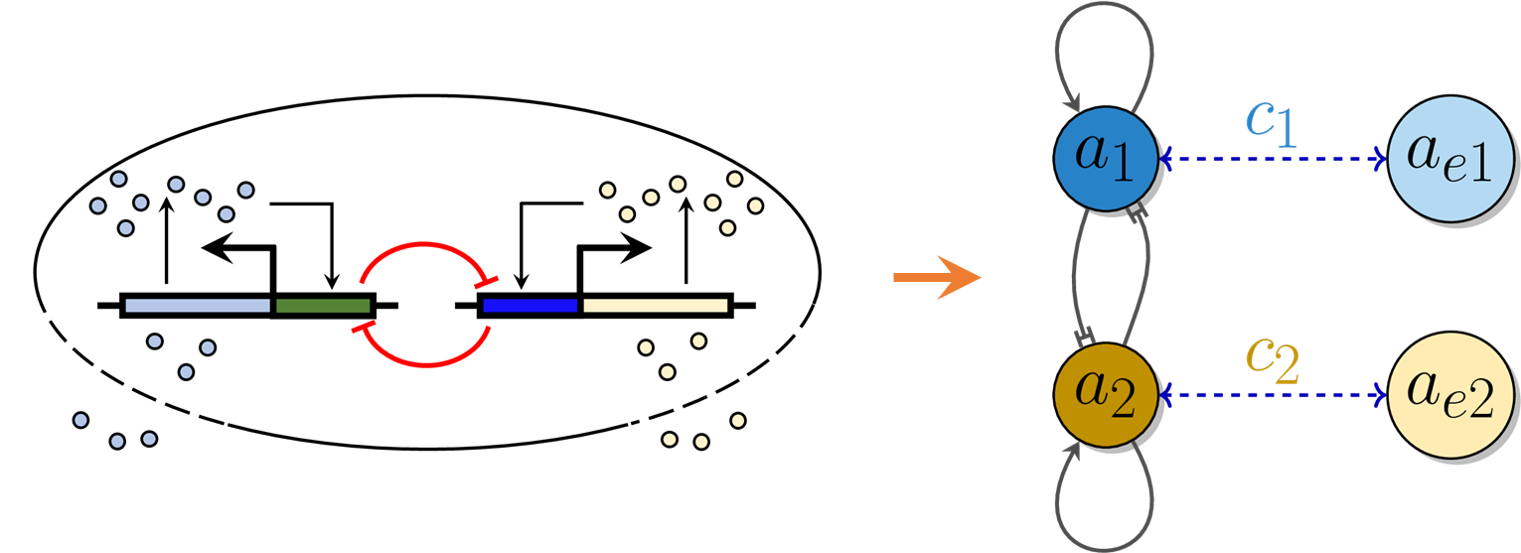}
\caption{Schematic of a bistable quorum-sensing toggle switch in a bacterial cell. The intracellular autoinducers $a_1$ and $a_2$ mutually repress each other's production while each self-activates its own expression. Each autoinducer exchanges with its extracellular pool ($a_{e1}$ or $a_{e2}$) at a rate controlled by the membrane permeability $c_1$ or $c_2$, respectively.}
\label{fig:toggle_switch}
\end{figure}

Each cell contains a quorum-sensing genetic toggle switch, illustrated in Fig.~\ref{fig:toggle_switch}. Two intracellular autoinducers, $a_1$ and $a_2$, self-activate their own production through Hill-type positive feedback and mutually repress each other. Each autoinducer exchanges with a shared extracellular pool ($a_{e1}$ or $a_{e2}$) at a rate controlled by the membrane permeability $c_1$ or $c_2$, respectively. 

The intracellular and extracellular dynamics are governed by:
\begin{subequations}\label{eq:qs_model}
\begin{align}
\frac{d a_{1}}{d t}
&=
\frac{\eta_{1}+k_{a1}\left(\frac{a_{1}}{k_{1}}\right)^{n}}
     {1+\left(\frac{a_{1}}{k_{1}}\right)^{n}}
+ \frac{k_{r1}}{1+\left(\frac{a_{2}}{k_{t1}}\right)^{n}}
\nonumber\\
&\quad
- d_{1} a_{1}
- c_{1}(a_{1}-a_{e1}),
\\[0.5em]
\frac{d a_{2}}{d t}
&=
\frac{\eta_{2}+k_{a2}\left(\frac{a_{2}}{k_{2}}\right)^{n}}
     {1+\left(\frac{a_{2}}{k_{2}}\right)^{n}}
+ \frac{k_{r2}}{1+\left(\frac{a_{1}}{k_{t2}}\right)^{n}}
\nonumber\\
&\quad
- d_{2} a_{2}
- c_{2}(a_{2}-a_{e2}),
\\[0.5em]
\frac{d a_{e1}}{d t}
&= \alpha_{1} c_{1}(a_{1}-a_{e1}),
\\[0.5em]
\frac{d a_{e2}}{d t}
&= \alpha_{2} c_{2}(a_{2}-a_{e2}).
\end{align}
\end{subequations}
Here, $k_1$ and $k_2$ denote the self-activation thresholds. The basal production rates of $a_1$ and $a_2$ are $\eta_1$ and $\eta_2$, respectively. $k_{a1}$ and $k_{a2}$ represent the maximum self-activation strengths for $a_1$ and $a_2$, respectively. The Hill exponent $n$ controls the steepness of all regulatory interactions. The cross-repression amplitudes $k_{r1}$ and $k_{r2}$ act on $a_1$ and $a_2$ respectively, with thresholds $k_{t1}$ and $k_{t2}$ setting the sensitivity of mutual inhibition. The intracellular autoinducers $a_1$ and $a_2$ degrade at rates $d_1$ and $d_2$, respectively. The permeability coupling term $c_k(a_k - a_{ek})$ drives the intracellular concentration toward the shared extracellular field, with secretion efficiencies $\alpha_1$ and $\alpha_2$. The extracellular signals $a_{e1}$ and $a_{e2}$ diffuse across the grid with equal coefficients $D_{e1} = D_{e2} \equiv D_{\rm e}$. All parameter values are given in Table~S1 of the Supplementary Material. The Supplementary Material will be made available with the published version.

\begin{figure}[ht]
\centering
\includegraphics[width=\linewidth]{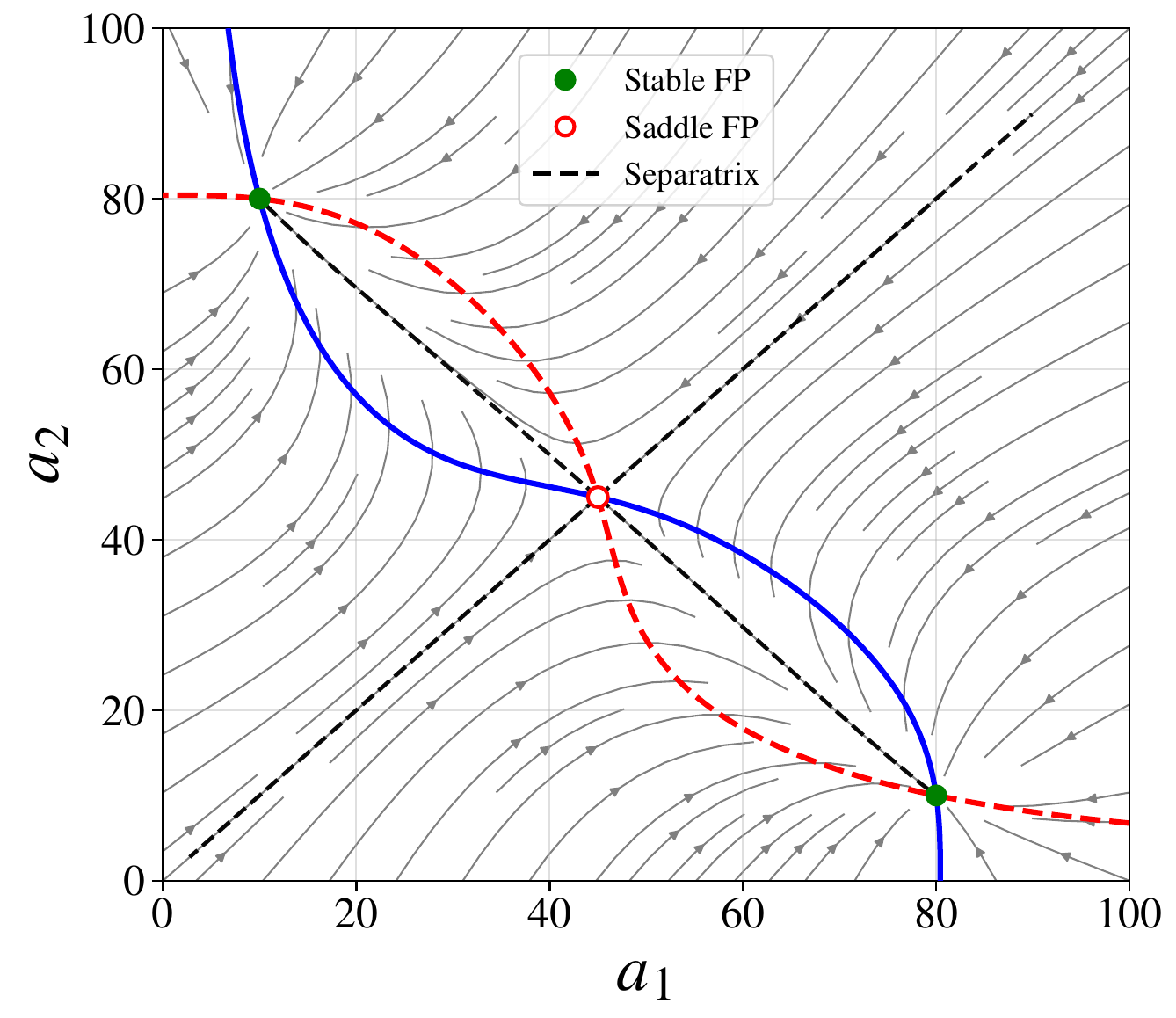}
\caption{Phase plane of the intracellular toggle-switch equations in 
$(a_1, a_2)$ space.
Gray arrows show the vector field, and trajectories converge to one of two stable fixed points (filled circles) at $(a_1, a_2) = (80, 10)$ and $(10, 80)$.
A saddle point (open circle) at $(a_1, a_2) = (45, 45)$ separates the two basins of attraction and acts as the decision boundary for permeability selection.}
\label{fig:phase_plane}
\end{figure}

The toggle-switch equations [Eqs.~\eqref{eq:qs_model}] exhibit bistability for the parameter values used throughout this study
(Table~S1).
Phase-plane analysis in the $(a_1, a_2)$ state space reveals two stable fixed
points, $(\bar a_{\rm ON}, \bar a_{\rm OFF})$ and
$(\bar a_{\rm OFF}, \bar a_{\rm ON})$, with $\bar a_{\rm ON} = 80$ and
$\bar a_{\rm OFF} = 10$. They are separated by a saddle point at
$(a_{\rm sad}, a_{\rm sad})$, with $a_{\rm sad} = 45$
(Fig.~\ref{fig:phase_plane}).
Trajectories starting above the saddle converge to the
high-$a_1$ state and those starting below converge to the
high-$a_2$ state.
The saddle point thus acts as the decision boundary: cells initialized on opposite sides of the saddle commit to mutually exclusive autoinducer states.
Bistability depends on two key parameters: the cross-repression threshold $k_t$ ($k_{t1}=k_{t2}$) and the self-activation threshold $k$ ($k_1=k_2$).
Bifurcation analysis reveals that both parameters produce
spontaneous symmetry breaking through a supercritical pitchfork bifurcation. Above a critical value of either parameter, the symmetric monostable state loses stability, and two stable asymmetric branches emerge.  The details are given in Sec.~I of the Supplementary Material. The parameter values used throughout this study, therefore, sit well within the bistable region.

\subsection{Population-level individual-based model}
\label{sec:IBM}
\begin{figure}[ht]
\centering
\includegraphics[width=\linewidth]{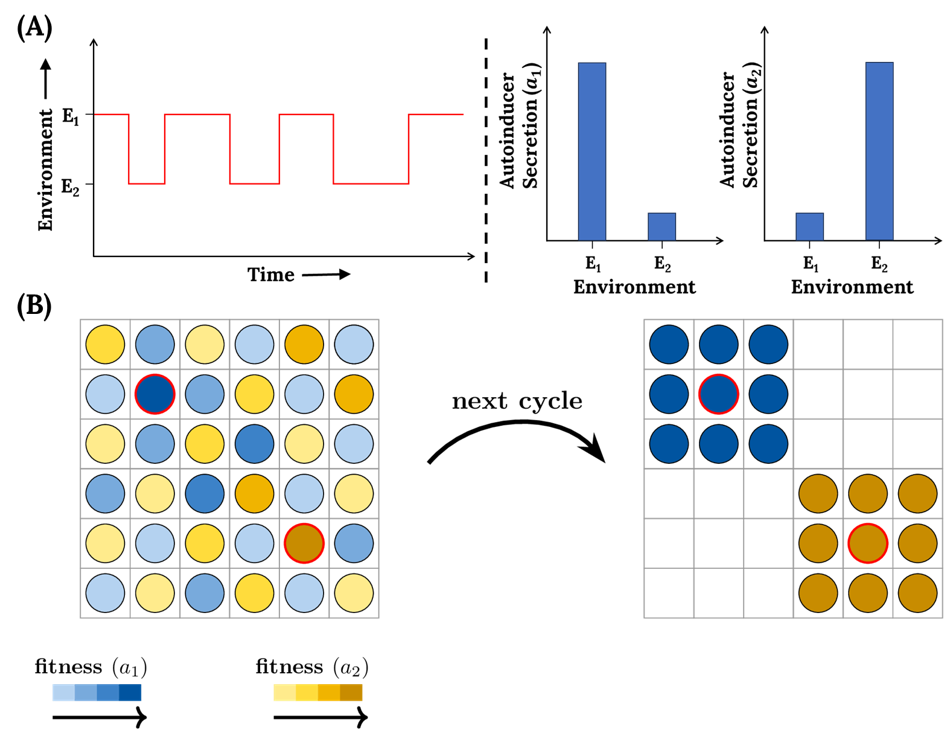}
\caption{Overview of the evolutionary simulation framework.
(A) The environment switches randomly each generation between
states $E_1$ and $E_2$, setting the target concentration for the
active autoinducer ($a_1$ in $E_1$, $a_2$ in $E_2$).
(B) Fitness-based selection after each generation:
in $E_1$, fitness depends on $a_1$; in $E_2$, it depends on $a_2$.}
\label{fig:envs}
\end{figure}

The cells, each carrying the toggle switch of Eqs.~\eqref{eq:qs_model}, sit
on a $50 \times 50$ periodic grid and communicate through the two shared
extracellular fields. Each generation proceeds in four steps
(Fig.~\ref{fig:envs}).

First, one environment is drawn at random: $E_1$ with probability $p_1$ or
$E_2$ with probability $p_2 = 1 - p_1$. In environment $E_k$, autoinducer
$a_k$ is the active signal and the other is inactive.

Second, the intracellular concentrations are reset. The active autoinducer
is drawn from a truncated normal distribution centered on the high fixed
point $\bar a_{\rm ON}$ with width $\sigma_{\rm ON}$, and the inactive one
around the low fixed point $\bar a_{\rm OFF}$ with width $\sigma_{\rm OFF}$.
Each cell therefore starts near the correct state, and the noise represents error in its commitment, not uncertainty about which environment is present. A cell
whose active concentration starts below the saddle $a_{\rm sad}$ heads for
the wrong attractor unless the shared field pulls it back. The extracellular
fields are not reset, so they carry information between generations.

Third, the toggle equations are integrated for all cells while the
extracellular fields diffuse across the grid by a Gaussian kernel of width
$\sigma_{\rm e} = \sqrt{2 D_{\rm e} \Delta t}$. Each cell's fitness then
rewards how closely its active autoinducer tracked $\bar a_{\rm ON}$ and
charges a metabolic cost for its total permeability $c_1 + c_2$. The
permeability of the inactive autoinducer also decays at rate $\rho$.

Fourth, offspring are chosen in proportion to fitness, keeping the
population size fixed. Each permeability mutates with probability $\mu$ by a
small step $\pm\sigma_{\rm mut}$, and offspring are placed near their parent.
The equations, the fitness function and all parameter values are given in Sec.~IX and Table~S2 of the Supplementary Material.

Each cell $n$ carries its own traits $c_i^{(n)}$. Unless stated otherwise, $c_i$ denotes the population mean in one trial, $c_i = N^{-1}\sum_{n} c_i^{(n)}$, and $\langle c_i \rangle$ its average over replicate trials.

\section{Results}
\label{sec:results}
We first show that permeability evolves through a collective extracellular mechanism at the population level, building on the single-cell bistability established in Sec.~\ref{sec:toggle}.
Next, we show how environmental frequency controls whether the subordinate trait evolves, through a smooth frequency crossover.
We then characterize how sensing noise controls the evolved
plateau through a non-monotonic optimum. Finally, we show that equal environmental frequencies produce stochastic coexistence, with noise asymmetry grading the level of coexistence rather than selecting a winner.
\subsection{Permeability evolves through collective sensing}
\label{sec:results_baseline}

\begin{figure*}[t]
  \centering
  \begin{subfigure}{\linewidth}
    \centering
    \includegraphics[width=\linewidth]{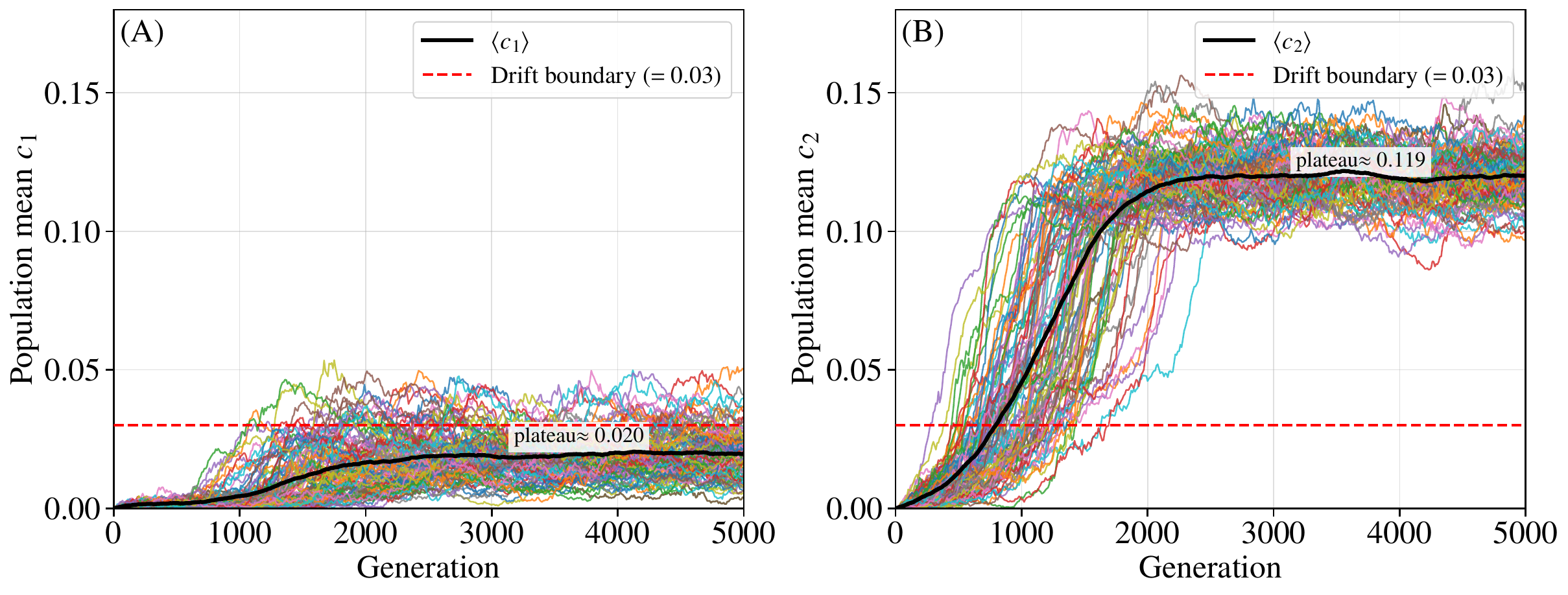}
  \end{subfigure}
  \vspace{0.6em} 
  \begin{subfigure}{\linewidth}
    \centering
    \includegraphics[width=\linewidth]{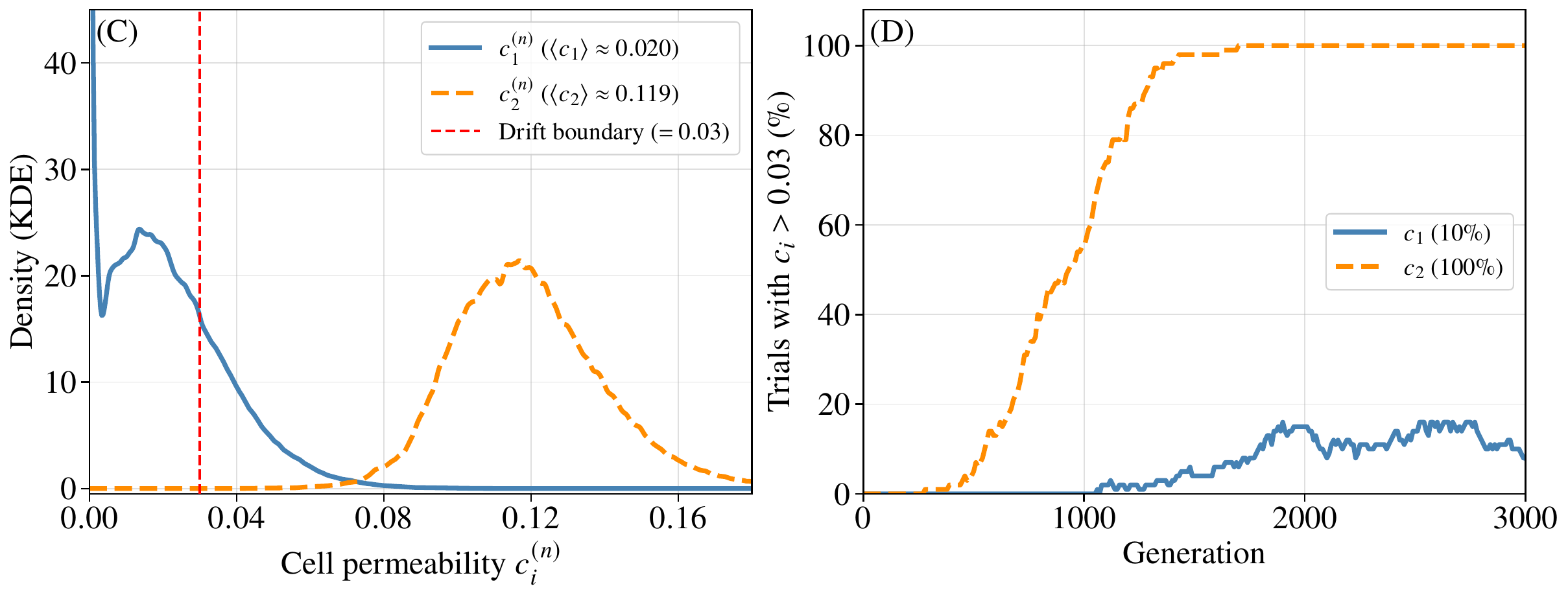}
  \end{subfigure}
\caption{Under strong environmental bias, only the dominant trait evolves, and the rare-environment trait is suppressed ($p_1 = 0.2$, $\sigma_{\rm ON} = 50$, $\sigma_{\rm OFF} = 20$, $N = 100$ trials).
(A,B)~Trajectories of $c_1$ and $c_2$. Colored curves are individual trials ($c_i$). The black curve is the mean over trials ($\langle c_i\rangle$). The dashed line marks the drift boundary ($0.03$). $\langle c_2\rangle$ reaches ${\approx}0.119$, while $\langle c_1\rangle$ stays near $0.020$.
(C)~Density of cell-level permeabilities $c_i^{(n)}$ at steady state.
(D)~Fraction of trials with $c_1$ (solid) and $c_2$ (dashed) above the drift boundary.}
\label{fig:evolution_c1c2_e2}
\end{figure*}

Fig.~\ref{fig:evolution_c1c2_e2} shows the evolutionary trajectories of $c_1$ and $c_2$ under asymmetric environmental pressure ($p_1 = 0.2$, $p_2 = 0.8$; $\sigma_{\rm ON} = 50$, $\sigma_{\rm OFF} = 20$; $N = 100$ trials). Under this environmental bias, $E_2$ occurs $80\%$ of the time and $a_2$ is the active signal in most generations. Consequently, the permeability $c_2$,  controlling $a_2$ secretion, experiences stronger and more consistent selection pressure than $c_1$.
As a result, $c_2$ evolves to a plateau of $\langle c_2\rangle \approx 0.119$, well above the drift boundary ($=0.03$), while $c_1$ stays below it throughout ($\langle c_1\rangle \approx 0.020$) (Figs.~\ref{fig:evolution_c1c2_e2}A,B). We take $c = 0.03$ as the drift boundary throughout, nearly three times the largest value attained by any unselected trait, so that a trait exceeding it cannot be explained by mutation and drift alone. The subordinate value is nonetheless not neutral: it exceeds the largest steady state reached by any of the $200$ unselected trials in the no-selection controls, so $c_1$ is suppressed rather than unselected.
We term this toggle-selective adaptation: only the permeability trait relevant to the dominant environment evolves to high levels, while the other is suppressed by stronger relaxation pressure ($p_2\rho = 0.0016$ per generation for $c_1$ vs $p_1\rho = 0.0004$ for $c_2$).

The cell-level distributions at steady state show the same asymmetry (Fig.~\ref{fig:evolution_c1c2_e2}C). The density of $c_1^{(n)}$ peaks near $0.02$, below the drift boundary. The density of $c_2^{(n)}$ is sharply peaked near $0.119$, so adaptation is shared across the population rather than carried by a few cells.
%
Individual trials adapt at very different times, spanning nearly $3000$
generations, yet all converge to the same plateau ${\approx}0.119$
(Fig.~\ref{fig:evolution_c1c2_e2}B).
Adaptation speed and plateau height are therefore controlled differently. The speed is set by the collective extracellular
buildup, while the plateau is set by the balance between the tracking benefit and the metabolic cost, with relaxation lowering it further, and independent of when evolution started.
The fraction of trials crossing the drift boundary confirms this picture (Fig.~\ref{fig:evolution_c1c2_e2}D): $c_2$ reaches $100\%$ of trials by generation $\sim\!1000$, while $c_1$ fluctuates near $10\%$ throughout, consistent with drift rather than selection.

\begin{figure*}[t]
  \centering
  \includegraphics[width=0.31\linewidth]{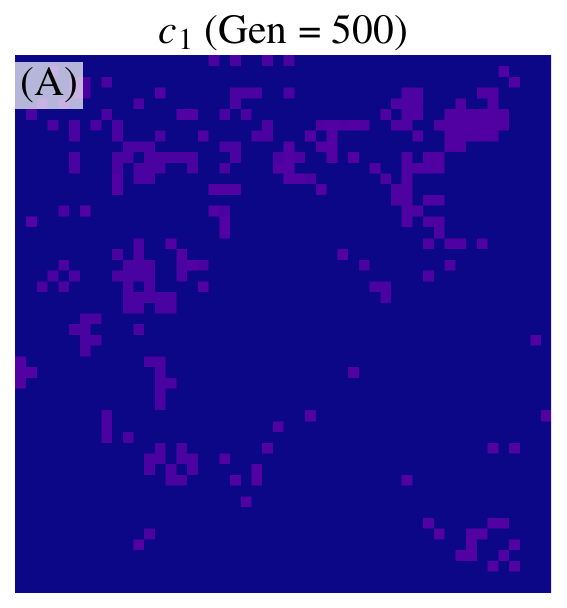}\hfill
  \includegraphics[width=0.31\linewidth]{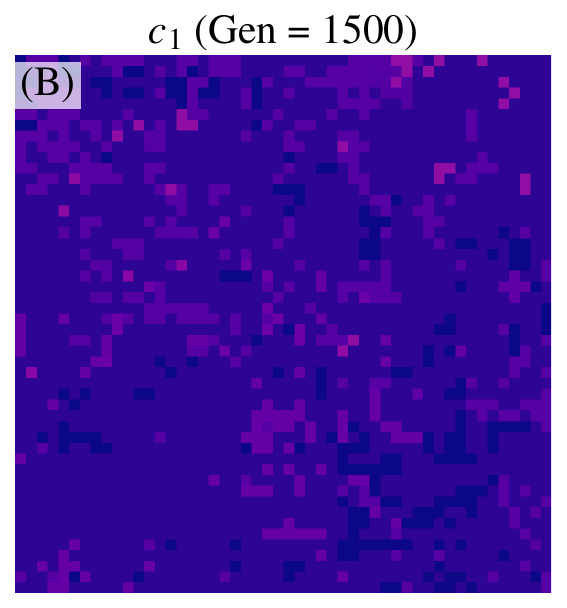}\hfill
  \includegraphics[width=0.38\linewidth]{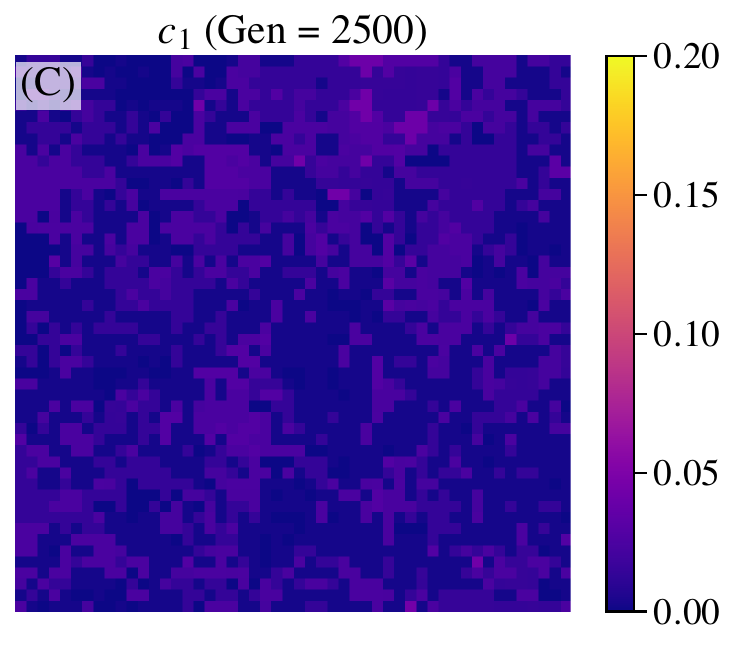}\\[0.6em]
  \includegraphics[width=0.31\linewidth]{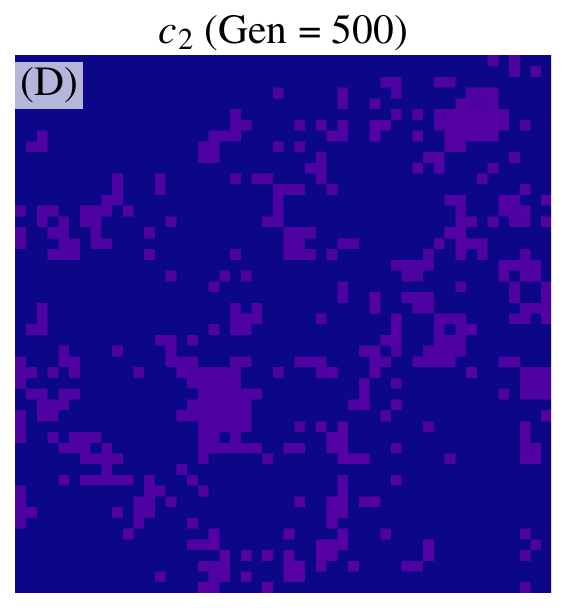}\hfill
  \includegraphics[width=0.31\linewidth]{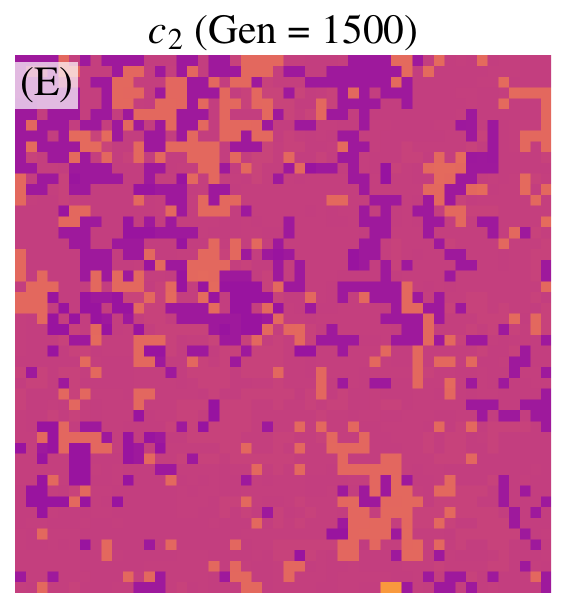}\hfill
  \includegraphics[width=0.38\linewidth]{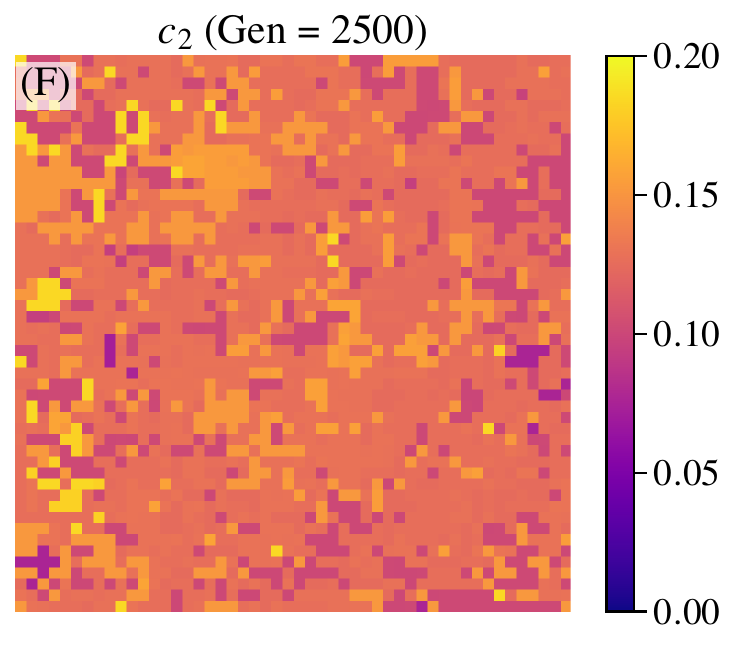}

\caption{High permeability spreads from local clusters, a signature of collective sensing. Spatial maps of permeability evolution on the $50\times50$ 
cell lattice under asymmetric environmental pressure 
($p_1=0.2$, $p_2=0.8$; $\sigma_{\rm ON}=50$, $\sigma_{\rm OFF}=20$).}
\label{fig:six_e1}
\end{figure*}

Fig.~\ref{fig:six_e1} shows spatial maps of $c_1$ and $c_2$ across the $50\times50$ cell grid at generations 500, 1500, and 2500. At generation
500, both traits are low and spatially heterogeneous. By generation 1500, $c_2$ has organized into high-value clusters while $c_1$ remains uniformly
low. By generation 2500, $c_2$ is high and spatially uniform across the grid, and $c_1$ is still near zero. The cluster stage points to a
collective mechanism. High-$c_2$ cells amplify the shared extracellular $a_{e2}$ field locally. Diffusion of $a_{e2}$ then rescues neighboring low-$c_2$ cells from the wrong attractor basin, and the high-permeability state spreads outward from the clusters.

The collective mechanism requires a shared extracellular field that reaches beyond a cell's immediate neighbors. We varied the diffusion coefficient $D_{\rm e}$ from $0$ to $8$ at $p_1 = 0.2$ and $\sigma_{\rm ON} = 50$
(Sec.~VII and Figs.~S6 and S8 of the Supplementary Material). Without diffusion, both traits stay at the mutation--drift floor ($\langle c_2\rangle = 0.008$). Each cell then senses only its own autoinducer, and the tracking gain does not repay the metabolic cost. The
response switches on sharply. Between $D_{\rm e} = 0.05$ and $0.1$, a diffusion range of three to five cell diameters, the dominant trait rises twelvefold, from $0.011$ to $0.128$. A longer range brings no further
benefit. At $D_{\rm e} = 8$, where the field is nearly uniform, the evolved permeability is about $10\%$ lower ($z = 4.9$, the difference in means over their combined standard error). Global mixing dilutes the local amplification that seeds high-permeability clusters
(Fig.~\ref{fig:six_e1}). Collective sensing in this model is therefore local.

A second control isolates the role of sensing noise itself. With \(\sigma_{\rm ON} = 1\) the active-channel estimate is effectively exact. Every cell initializes at its correct target and commits to the appropriate autoinducer state, so tracking accuracy no longer depends on permeability and all cells receive the same fitness. With no fitness variation there is no selection differential, and neither \(c_1\) nor \(c_2\) evolves (Fig.~S5). Sensing noise is what generates the mis-committed cells, and their rescue by the shared autoinducer field is what makes higher permeability advantageous.
\subsection{Environmental frequency controls subordinate-trait evolution}
\label{sec:results_threshold}

To determine how environmental frequency shapes permeability evolution, we
measured the steady-state mean permeabilities $\langle c_1 \rangle$ and
$\langle c_2 \rangle$ across $p_1 \in [0,1]$ ($\sigma_{\rm ON} = 50$,
$\sigma_{\rm OFF} = 20$; $N = 100$ trials per condition), averaging over the
last $1000$ generations of each run.

\begin{figure*}[t]
  \centering
  \includegraphics[width=\linewidth]{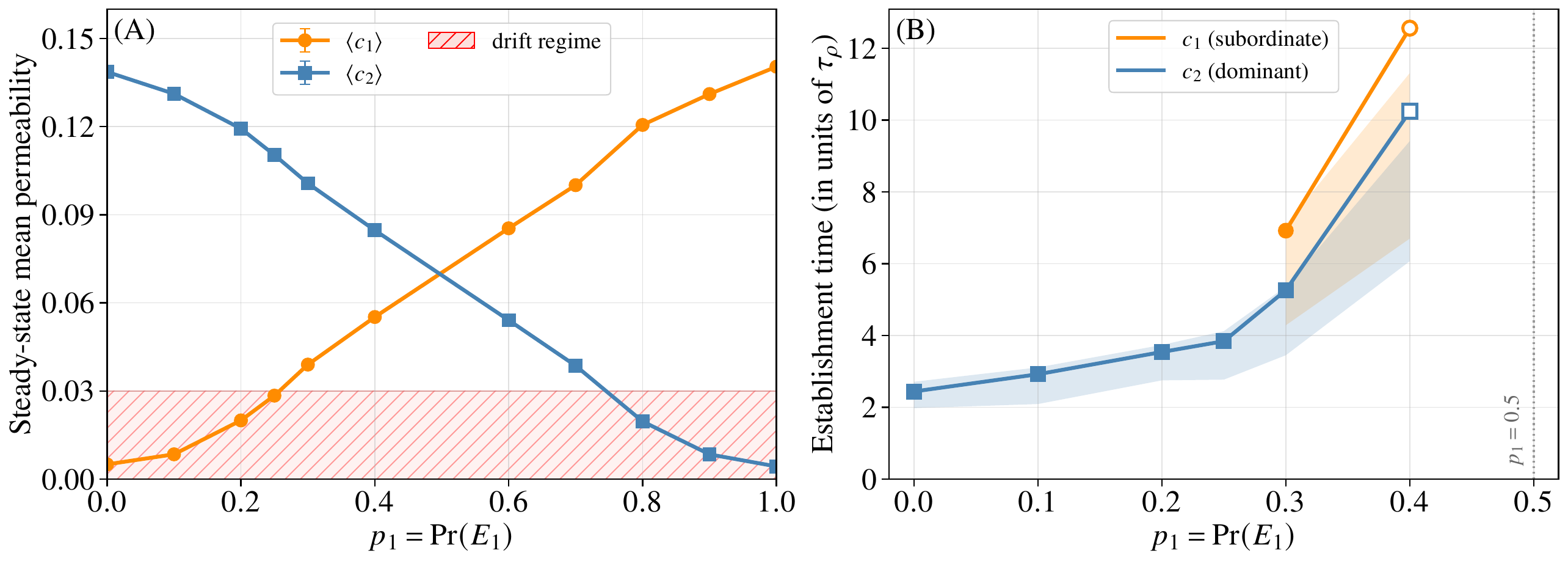}
\caption{As environments become equally frequent, the two plateaus converge while the time to reach them lengthens ($\sigma_{\rm ON} = 50$, $\sigma_{\rm OFF} = 20$, $N = 91$--$100$ trials per point).
(A)~Steady-state mean permeabilities $\langle c_1\rangle$ (circles) and $\langle c_2\rangle$ (squares) against $p_1$, averaged over the last $1000$ generations. The pink band marks the drift regime ($c_i < 0.03$). Error bars are $\pm 1$ SEM and mostly smaller than the markers.
(B)~Establishment time of each trait in units of $\tau_\rho = 1/\rho = 500$ generations. The line marks when the population-mean trajectory reaches $90\%$ of its plateau. The shaded band is the interquartile range across individual trials. Only $p_1 \le 0.4$ is shown because the outcome is symmetric under $p_1 \leftrightarrow 1-p_1$. Open markers at $p_1 = 0.4$ are lower bounds, since curves are still rising at the end of the run.
$p_1 = 0.5$ is omitted from both panels because only about $40\%$ of trials resolve there.}
\label{fig:steady_state_phi}
\end{figure*}

Fig.~\ref{fig:steady_state_phi}A marks the drift boundary at $c_i = 0.03$
(pink band), below which a trait sits near the mutation--drift baseline.
As $p_1$ increases, $\langle c_1 \rangle$ rises monotonically while $\langle c_2 \rangle$ falls, the two crossing at $p_1 = 0.5$. The subordinate trait stays low when its
environment is rare and rises smoothly as that environment becomes more
frequent, clearing the drift boundary just above $p_1 = 0.25$
($\langle c_1\rangle = 0.028$ at $p_1 = 0.25$ and $0.039$ at $p_1 = 0.30$). The
onset is a crossover rather than a threshold, and it is symmetric under
$p_1 \leftrightarrow 1-p_1$. In the deterministic limit it sharpens into a
transcritical bifurcation that the mutation--drift floor unfolds
(Sec.~\ref{sec:theory}).

The two ends of this range behave differently. At the extremes, suppression is
toggle-selective: at $p_1 = 0.1$ the subordinate
$\langle c_1 \rangle \approx 0.008$ is barely distinguishable from the neutral
level while the dominant $\langle c_2 \rangle \approx 0.131$ is high, with roles
reversed at $p_1 = 0.9$. Toward the center both traits are well above the drift boundary: at
$p_1 = 0.4$, $\langle c_2 \rangle \approx 0.085$ leads but
$\langle c_1 \rangle \approx 0.055$ is well above the boundary, with the mirror
at $p_1 = 0.6$. The asymmetry arises from differential relaxation pressure: the locus inactive in the dominant environment decays
$\max(p_1, 1-p_1)/\min(p_1, 1-p_1)$ times faster than the active locus, holding it nearer the drift floor the further $p_1$ is from $0.5$.

Level and timing are separately controlled, and across the frequency range they run in opposite directions. Fig.~\ref{fig:steady_state_phi}B gives the establishment time of each trait: the first generation at which the population-mean trajectory reaches $90\%$ of its own plateau, in units of the relaxation time $\tau_\rho = 1/\rho = 500$ generations. Because the outcome is symmetric under $p_1 \leftrightarrow 1-p_1$ with the two traits exchanged, only $p_1 \le 0.4$ is shown. We refer the criterion to the population-mean trajectory, not to a common threshold, because individual trials plateau at widely different levels (coefficient of variation $0.09$ for the dominant trait and $0.21$ for the subordinate at $p_1 = 0.3$), so a fixed target would confound when a trial adapted with how far it went. The ordering and the trend are
unchanged for targets between $80\%$ and $99\%$ of the plateau.

The dominant trait establishes in $2.4\,\tau_\rho$ when its environment is the only one that occurs. It establishes progressively later as the second environment becomes more common, taking $3.5\,\tau_\rho$ at $p_1 = 0.2$ and $10.2\,\tau_\rho$ at $p_1 = 0.4$, a more than fourfold slowdown. The subordinate appears once its plateau clears the drift boundary, at $p_1 = 0.3$. It takes about a quarter longer than the dominant trait, $6.9$ against $5.3\,\tau_\rho$ at $p_1 = 0.3$ and $12.6$ against $10.2\,\tau_\rho$ at $p_1 = 0.4$. The two traits differ far more in the level they reach than in the time they take to reach it.
 At $p_1 = 0.4$ the mean curves are still rising at generation $10\,000$, so the times there are lower bounds, and at $p_1 = 0.5$ about $60\%$ of trials remain unresolved entirely (Sec.~\ref{sec:results_bistability}). As the two environments approach equal frequency the two plateaus converge while the approach to them lengthens: the population ends up maintaining both channels at comparable levels, and takes progressively longer to get there.

\subsection{Sensing noise has a non-monotonic effect}
\label{sec:results_noise}

\begin{figure*}[htbp]
\centering
\includegraphics[width=\linewidth]{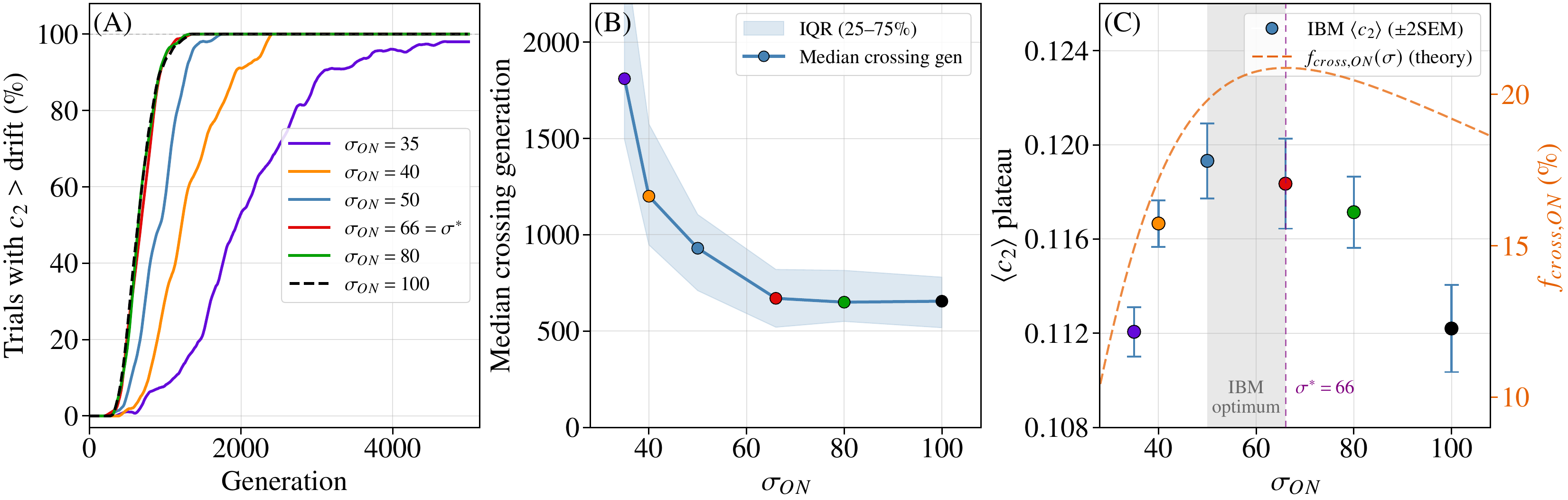}
\caption{Sensing noise speeds adaptation monotonically but has an interior optimum for the evolved plateau ($p_1 = 0.2$, $\sigma_{\rm OFF} = 20$, $N = 100$ trials per $\sigma_{\rm ON}$).
(A)~Fraction of trials with $c_2$ above the drift boundary ($0.03$) for six values of $\sigma_{\rm ON}$.
(B)~Median generation at which trials cross the drift boundary. The shaded band is the interquartile range.
(C)~Mean plateau $\langle c_2 \rangle$ ($\pm 2$ SEM) against $\sigma_{\rm ON}$, with the saddle-crossing fraction $f_{\rm cross,ON}$ (dashed, right axis). The grey band marks $\sigma_{\rm ON} \approx 50$--$66$, where the IBM maximum lies. The vertical line marks the model optimum $\sigma^{*} \approx 66$.}
\label{fig:noise_effect}
\end{figure*}

Sensing noise is a precondition for permeability to evolve at all (Sec.~\ref{sec:results_baseline}). We next asked how much of it is optimal. We varied $\sigma_{\rm ON} \in \{35, 40, 50, 66, 80, 100\}$ while holding all other parameters fixed ($p_1 = 0.2$, $\sigma_{\rm OFF} = 20$; $N = 100$ trials per condition).
All conditions eventually reach $100\%$ adaptation, but higher noise accelerates the onset (Fig.~\ref{fig:noise_effect}A). At $\sigma_{\rm ON} = 35$ the population needs ${\sim}4000$ generations to fully adapt, whereas the highest-noise conditions converge by generation ${\sim}1000$. The median crossing generation falls with $\sigma_{\rm ON}$ and levels off beyond $\sigma_{\rm ON} \approx 66$ (Fig.~\ref{fig:noise_effect}B). Higher noise places more cells below the
toggle saddle each generation, which seeds the collective extracellular buildup sooner. Once enough cells are seeded, speed approaches a noise-independent limit.

The plateau height behaves differently. It is non-monotonic in $\sigma_{\rm ON}$ (Fig.~\ref{fig:noise_effect}C). The mean steady-state $\langle c_2 \rangle$ rises from $0.1121$ at $\sigma_{\rm ON} = 35$ to $0.1193$ at $\sigma_{\rm ON} = 50$, then falls to $0.1122$ at $\sigma_{\rm ON} = 100$. Both ends lie more than five standard errors below $\sigma_{\rm ON} = 50$ ($z = 7.6$ and $5.9$). Intermediate sensing noise is therefore optimal for permeability evolution.

The location of the optimum is less sharply determined than its existence. The plateaus at $\sigma_{\rm ON} = 50$ and $66$ differ by $0.0010$ against a combined standard error of $0.0013$ ($z = 0.78$). The maximum is therefore resolved only to the interval $\sigma_{\rm ON} \approx 50$--$66$ (pairwise comparisons in Sec.~VI of the Supplementary Material). The dashed curve in Fig.~\ref{fig:noise_effect}C shows the saddle-crossing fraction $f_{\rm cross,ON}(\sigma_{\rm ON})$ (Sec.~X of the Supplementary Material). It peaks at $\sigma^{*} \approx 66$, inside the interval the data allow, and follows from
the toggle geometry alone. At low noise few cells fall below the saddle. At
high noise, truncation at zero pushes the extra spread above the active
state, and fewer cells fall below it again.
The subordinate trait $\langle c_1 \rangle$ follows the same low-noise trend
but does not decline at high noise. It rises from $0.0131$ at $\sigma_{\rm ON}=35$ to $0.0236$ at $66$, then stays level up to $100$ ($0.0225$ and $0.0229$; $z<1.1$). Its optimum is therefore $\sigma_{\rm ON}\gtrsim 66$, consistent with $\sigma^{*}$.
Sensing noise thus affects adaptation speed and plateau height differently: speed rises and saturates, while the plateau has an interior optimum.

The results above hold the inactive-channel noise at $\sigma_{\rm OFF} = 20$. Because this value is stipulated, we varied it over $[1,60]$ (Sec.~V of the
Supplementary Material). The dominant trait changes little above the baseline, but the subordinate trait rises fifteenfold and crosses the drift
boundary near $\sigma_{\rm OFF} \approx 30$. Toggle-selective adaptation at $p_1 = 0.2$ therefore requires low inactive-channel noise as well as
environmental asymmetry.
\subsection{Symmetric environments produce stochastic coexistence}
\label{sec:results_bistability}

\begin{figure*}[htbp]
\centering
\includegraphics[width=\linewidth]{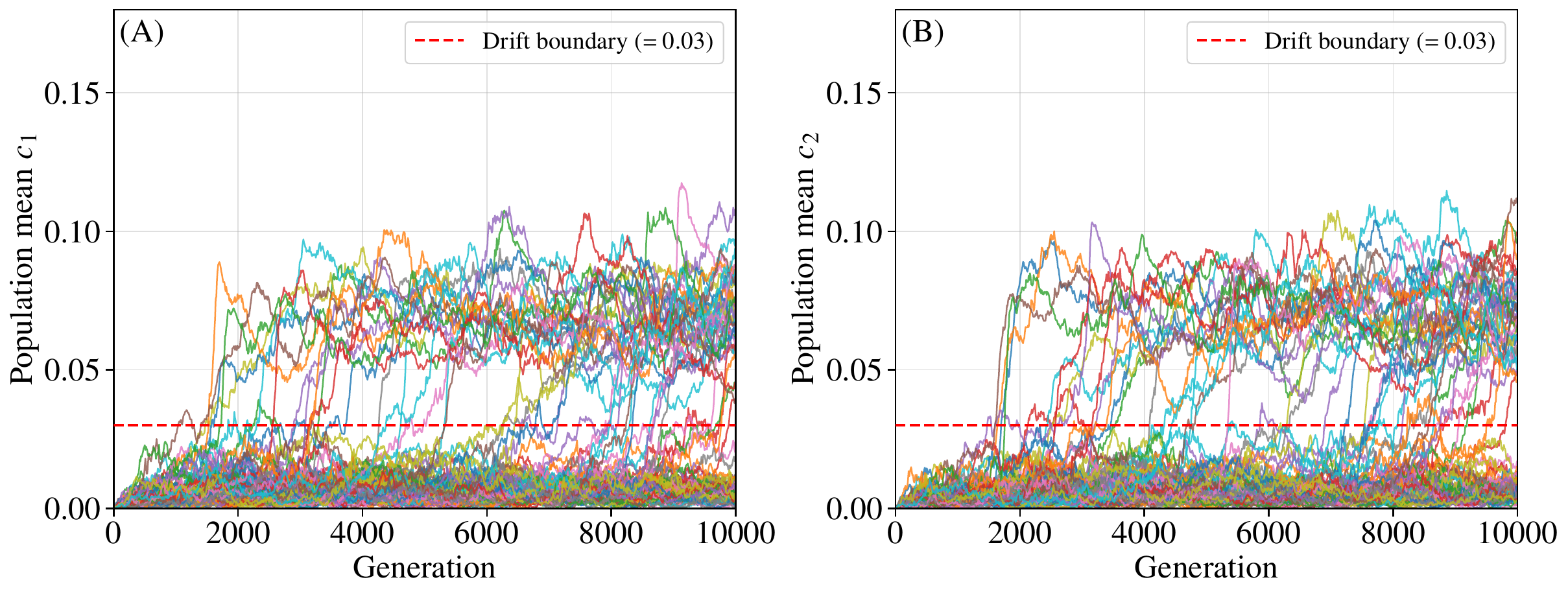}
\caption{At equal frequencies, adaptation is slow and stochastic in timing ($p_1 = 0.5$, $\sigma_{\rm ON}^{(1)} = \sigma_{\rm ON}^{(2)} = 50$, $\sigma_{\rm OFF} = 20$, $N = 100$ trials).
(A,B)~Trajectories of $c_1$ and $c_2$. Colored curves are individual trials ($c_i$). The dashed line marks the drift boundary ($0.03$). The joint outcome is shown in Fig.~\ref{fig:e05_traj}.}
\label{fig:e05}
\end{figure*}

When both environments occur with equal probability ($p_1 = p_2 = 0.5$), the
selective pressure on $c_1$ and $c_2$ is symmetric by construction.
Fig.~\ref{fig:e05} shows the raw trajectories under symmetric sensing noise,
where the two active channels carry equal noise
$\sigma_{\rm ON}^{(1)} = \sigma_{\rm ON}^{(2)} = 50$ ($N = 100$ trials). Here
$\sigma_{\rm ON}^{(k)}$ is the active-channel noise of autoinducer $a_k$;
elsewhere the two are matched and written $\sigma_{\rm ON}$. Both traits evolve
slowly and stochastically. By generation $10{,}000$, only ${\sim}40\%$ of trials
have a given permeability cross the drift boundary. Within a trial, one trait
typically crosses first, while the other remains near the drift floor. This initially gives the appearance of winner-take-all dynamics. However, the lagging trait usually rises later: among trials that begin with a single leading trait, $83\%$ eventually have the second trait also cross the drift boundary. The stochasticity lies in the timing of adaptation, not in the identity of a winner.

\begin{figure*}[htbp]
\centering
\includegraphics[width=0.95\linewidth]{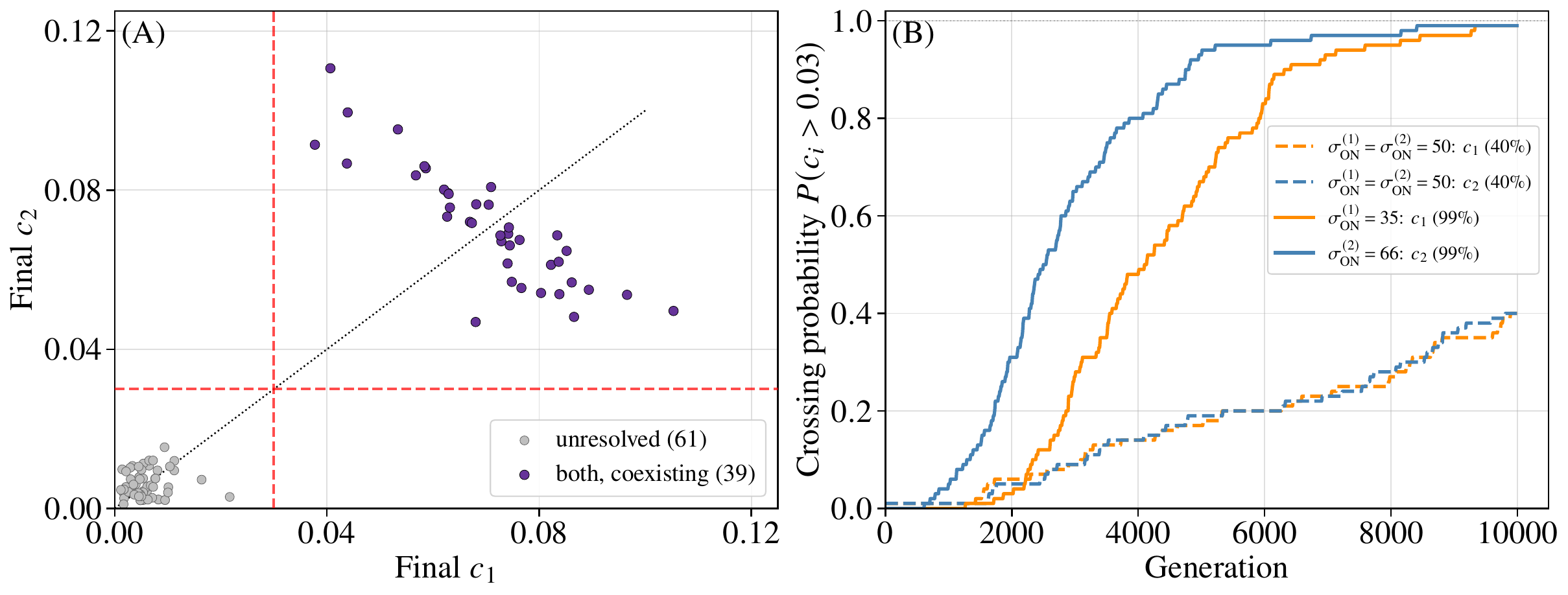}
\caption{At equal frequencies both traits coexist, and noise asymmetry grades the coexistence rather than selecting a winner ($p_1 = 0.5$, $\sigma_{\rm OFF} = 20$, $N = 100$ trials).
(A)~Final $(c_1, c_2)$ of each trial under symmetric noise ($\sigma_{\rm ON}^{(1)} = \sigma_{\rm ON}^{(2)} = 50$). Dashed lines mark the drift boundary ($0.03$). Purple points ($39$) have both traits elevated, and grey points ($61$) stay near the floor. No trial ends with a single trait elevated.
(B)~Fraction of trials with each trait above the drift boundary, against generation. Dashed curves are symmetric noise ($50/50$) and solid curves are asymmetric noise ($35/66$).}
\label{fig:e05_traj}
\end{figure*}

The joint endpoint distribution confirms coexistence (Fig.~\ref{fig:e05_traj}A). Resolved trials cluster around the $c_1 = c_2$ diagonal. A mean on the diagonal could also arise from two symmetric
off-diagonal attractors, but the split $\delta = c_1 - c_2$ shows no dip at zero (Fig.~S9 of the Supplementary Material). The trials, therefore, converge to a single coexistence state.
The two directions are very differently constrained. The total $c_1 + c_2$ is held at $0.1423$ with a standard deviation of $0.0052$, while the split $\delta$ has a standard deviation of $0.0206$, four times larger. Selection at the symmetric point therefore fixes how much export capacity the population maintains in total, and is nearly indifferent to how that capacity is divided between the
two channels. The attractor is a single node with a soft direction along the anti-diagonal, not a pair of competing states.
The adaptive-dynamics model is consistent with this reading. With weak cross-repression it has a single stable fixed point at $p_1 = 0.5$, on the diagonal at $(c_1, c_2) = (0.072, 0.072)$ (Sec.~\ref{sec:model_bistable}). The repression strength is not determined by the data (Sec.~XI of the Supplementary Material), so the model alone does not establish the single-node structure. The $\delta$ distribution in Fig.~S9 does.

Sensing-noise asymmetry does not convert coexistence into fixation. Instead, it
biases the level at which the two traits coexist, as shown in Fig.~\ref{fig:e05_traj}B, which compares the per-trait crossing probability for symmetric and asymmetric noise. The probability of crossing is the fraction of trials in which a trait crosses the drift boundary and stays above it. For symmetric noise we set $\sigma_{\rm ON}^{(1)} = \sigma_{\rm ON}^{(2)} = 50$; for
asymmetric noise, $\sigma_{\rm ON}^{(1)} = 35$ and $\sigma_{\rm ON}^{(2)} = 66$.
Under symmetric noise the two curves overlap and rise slowly, reaching only about
$40\%$ by generation $10{,}000$. Under asymmetric noise both traits adapt far
more reliably, each crossing in $99\%$ of trials, and the channel closer to the
saddle-crossing optimum leads: $c_2$, with $\sigma_{\rm ON}^{(2)} = 66 \approx \sigma^*$, crosses roughly $2000$ generations before $c_1$.
The asymmetry persists in the final levels. The optimal-noise channel settles higher, $\langle c_2\rangle \approx 0.074$ against $\langle c_1\rangle \approx 0.060$, and $c_2 > c_1$ in $82\%$ of resolved trials. Yet both traits still coexist. Noise asymmetry therefore biases the coexistence level without selecting a winner. The model reproduces this shift (Sec.~\ref{sec:model_bistable}).

\begin{figure*}[t]
\centering
\includegraphics[width=\linewidth]{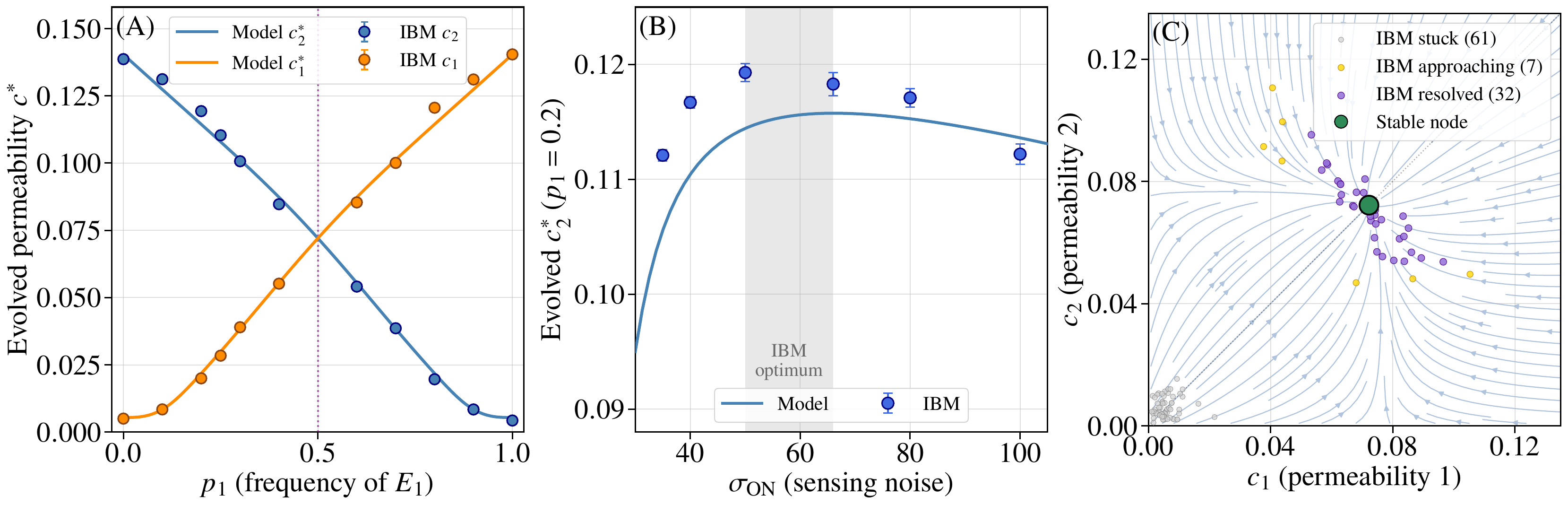}
\caption{One adaptive-dynamics model, with a single parameter set, reproduces both the asymmetric steady states and the single coexistence node at symmetry.
(A)~Steady-state permeability against $p_1$ ($\sigma_{\rm ON} = 50$, $\sigma_{\rm OFF} = 20$). Circles are IBM means $\pm$ SEM and curves are the model ($c_2$ blue, $c_1$ orange, RMSE $= 0.0023$, $R^2 = 0.998$).
(B)~Noise dependence of $c_2^{*}$ at $p_1 = 0.2$. The model was fitted at $\sigma_{\rm ON} = 50$ only, so this panel is a prediction. The grey band marks the IBM maximum ($\sigma_{\rm ON} \approx 50$--$66$).
(C)~Phase plane at $p_1 = 0.5$. Streamlines and the stable node (green) at $c_1^{*} = c_2^{*} = 0.072$ are model predictions. Points are IBM trials at generation $10\,000$, near the floor (grey, $61$), with one trait between $0.03$ and $0.05$ (gold, $7$), and with both traits above $0.05$ (purple, $32$).}
\label{fig:unified_2d}
\end{figure*}
\section{Adaptive-dynamics model of permeability evolution}
\label{sec:theory}

Having established the evolutionary outcomes empirically
(Section~\ref{sec:results}), we now develop an adaptive-dynamics framework that accounts for them. The framework is built on a single two-trait deterministic model. 


\subsection{Two-trait coupled dynamics}
\label{sec:model_2d}

We describe the joint evolution of the two membrane-permeability traits $c_1$
and $c_2$. The environment alternates between two states, $E_1$ and $E_2$. These states occur with frequencies $p_1$ and $p_2 = 1 - p_1$ and state $E_i$ favors trait $c_i$.
Each trait evolves under three forces. First, selection pushes it toward higher permeability. Second, relaxation pulls it toward the drift floor $b$ in the absence of its environment. In the IBM relaxation acts toward zero. The floor $b$ is the small level that upward mutation maintains against it, since mutation cannot push $c_i$ below zero. Third, cross-repression from the toggle switch acts on it.

The coupled adaptive dynamics are
\begin{align}
  \dot c_1 &= S_1\Bigl(1-{c_1}/{c_{\max}}\Bigr) c_1\, R(c_2)
              - p_2\,\kappa\,(c_1 - b), \label{eq:dc1}\\[2pt]
  \dot c_2 &= S_2\Bigl(1-{c_2}/{c_{\max}}\Bigr) c_2\, R(c_1)
              - p_1\,\kappa\,(c_2 - b), \label{eq:dc2}
\end{align}
where $S_i$ is the selection strength on trait $i$, $R(x)$ the cross-repression factor, $c_{\max}$ the permeability ceiling and $\kappa$ the relaxation rate. $S_i$ and $R(x)$ are given in Eqs.~\eqref{eq:selection} and \eqref{eq:repression} below.
The relaxation term is the coarse-grained counterpart of the per-generation decay of the inactive locus in the individual-based model (Sec.~IX of the Supplementary Material). It also absorbs the constant metabolic cost charged on the inactive locus. The relaxation of $c_2$ scales with $p_1$, the frequency of the environment in which $c_2$ is \emph{not} favored, and symmetrically for $c_1$.

The selection strength on trait $i$,
\begin{equation}
  S_i = \lambda_0\,
        f_{\rm cross}\bigl(\sigma_{\rm ON}^{(i)}\bigr)\,
        \frac{p_i^{\,h}}{p_i^{\,h} + K^{h}},
  \label{eq:selection}
\end{equation}
is a Hill function of the frequency $p_i$ of the environment favoring that trait (coefficient $h$, half-saturation $K$, $p_1$ and $p_2 = 1-p_1$),
scaled by the maximum selection strength $\lambda_0$ and by the active-channel noise factor $f_{\rm cross}(\sigma_{\rm ON}^{(i)})$. The noise factor is evaluated at that trait's own active sensing noise $\sigma_{\rm ON}^{(i)}$ (Sec.~X of the Supplementary Material). It tunes the strength of selection, and carries the non-monotonic noise optimum (Sec.~\ref{sec:model_noise}).

The cross-repression factor
\begin{equation}
  R(x) = \frac{1}{1 + (x/c_{\rm rep})^{m}}
  \label{eq:repression}
\end{equation}
$R(0)=1$, so a trait feels its full selection when the competing trait is absent, and $R(c_{\rm rep})=\tfrac12$, so $c_{\rm rep}$ is the competing level that halves selection. The form is phenomenological. In the intracellular model, cross-repression acts between autoinducer concentrations [Eqs.~\eqref{eq:qs_model}]. Here it acts between population traits. A population with high $c_j$ couples its cells strongly to the $a_{ej}$ field and pulls them toward the $a_j$ state, which reduces the benefit of raising $c_i$. Eq.~\eqref{eq:repression} captures this effect but is not derived from the cell equations. 

The steady states of Eqs.~\eqref{eq:dc1}--\eqref{eq:dc2} depend on $\lambda_0$ and $\kappa$ only through their ratio. Setting $\dot c_i = 0$ and dividing by $\kappa$ leaves $S_i/\kappa$ as the only combination that appears. A fit to steady-state data therefore determines $\lambda_0/\kappa$ and $K$ but leaves the overall timescale free. We fix it by setting $\kappa$ equal to the microscopic relaxation rate $\rho = 0.002$ per generation, so that model time is measured in generations and the dynamics, not only the fixed points, can be compared with the IBM.

In the frequency and noise sweeps, both active channels carry the same noise:
$\sigma_{\rm ON}^{(1)} = \sigma_{\rm ON}^{(2)} = \sigma_{\rm ON}$. So Eq.~\eqref{eq:selection}
reduces to a single $f_{\rm cross}(\sigma_{\rm ON})$. The symmetric-point
analysis of noise asymmetry (Sec.~\ref{sec:model_bistable}) is different. There,
we take $\sigma_{\rm ON}^{(1)} \neq \sigma_{\rm ON}^{(2)}$.
The fit is good as shown in Fig.~\ref{fig:unified_2d}A ($R^2 = 0.998$, RMSE $= 0.0023$). All model parameters are listed in Table~S3 of the Supplementary Material.

Fixing the time unit turns the adaptation dynamics into a prediction, since no
timing data entered the fit. Integrating
Eqs.~\eqref{eq:dc1}--\eqref{eq:dc2} from the drift floor and recording when the
dominant trait crosses $c = 0.03$ gives $1054$ generations at $p_1 = 0.2$,
against a median of ${\sim}950$ in the IBM, and $1139$ at $p_1 = 0.3$ and $0.7$
against $1400$ and $1300$. Away from symmetry the mean-field description
therefore reproduces the adaptation time to within about $20\%$.

Nearer symmetry it does not. At $p_1 = 0.4$ and $0.6$ the model predicts $1244$
generations while the IBM takes $3000$ and $3250$, a shortfall of a factor of
$2.5$ in the same direction at both frequencies. The discrepancy is not a
failure of the steady states, which the model reproduces at these frequencies to
within $2\times10^{-3}$, but of the approach to them. In the IBM, adaptation requires a local
cluster of high-permeability cells to grow large enough for the shared
extracellular field to sustain it (Fig.~\ref{fig:six_e1}), and the waiting time
for such a cluster lengthens as the selection differential between the two
traits weakens. Eqs.~\eqref{eq:dc1}--\eqref{eq:dc2} have positive drift
everywhere above the floor and so admit no such waiting time. The same omission
accounts for the trials that remain at the mutation--drift floor at $p_1 = 0.5$
(Sec.~\ref{sec:results_bistability}), which the deterministic flow cannot
produce.

The inactive-channel noise $\sigma_{\rm OFF}$ does not appear explicitly.
Occasional firing of the inactive autoinducer across the toggle saddle is a necessary condition for any permeability to evolve (Sec.~X of the Supplementary Material), but $\sigma_{\rm OFF}$ is held at the baseline value $20$ in every result reported in the main text, so its effect is constant and is absorbed into $\lambda_0$. Its influence on the evolved permeability is examined separately in Sec.~V of the Supplementary Material.
\subsection{The noise optimum}
\label{sec:model_noise}

The saddle-crossing fraction $f_{\rm cross}(\sigma)$ inherits a non-monotonic dependence on $\sigma_{\rm ON}$ from the truncated-normal geometry. At $\sigma \to 0$ the distribution concentrates at $\bar a_{\rm ON} = 80$ and no cell reaches the saddle at $45$; as $\sigma \to \infty$ truncation at zero removes mass from the active state and the crossing fraction collapses again.
Both limits vanish, so an interior maximum must exist. It occurs at
$\sigma^{*} \approx 66$, set by the separation between the active fixed point and the saddle. Through Eq.~\eqref{eq:selection} this non-monotonicity is inherited by the selection strength and hence by the evolved plateau. The model captures the low-noise side of both traits closely (Sec.~\ref{sec:results_noise}). Above the optimum, it predicts a shallow decline for both. The IBM dominant trait declines more steeply, while the subordinate trait does not decline.

Fig.~\ref{fig:unified_2d}B compares the resulting prediction with the IBM. The core parameters $\lambda_0$, $\kappa$ and $K$ were fitted to the $p_1$-sweep at fixed $\sigma_{\rm ON} = 50$ [Sec.~\ref{sec:model_2d}], so the noise dependence is an out-of-sample prediction rather than a fit. The model reproduces the interior maximum and places it at $\sigma^{*} = 66$, within the $\sigma_{\rm ON} \approx 50$--$66$ interval that the IBM data resolve (Sec.~\ref{sec:results_noise}). The agreement is qualitative: the predicted curve falls more steeply at low noise than the IBM does, reaching $0.106$ at $\sigma_{\rm ON} = 35$ against the measured $0.112$. The model therefore overstates how sensitively the evolved plateau tracks $f_{\rm cross}$, while capturing the geometric origin of the optimum.

\subsection{Reduction to the one-trait model}
\label{sec:model_1d}

Away from $p_1 = 0.5$, the two traits separate by timescale. Consider
$p_1 < 0.5$, where $c_2$ is favored, and the disfavored trait $c_1$ then
experiences little selection. It settles to a small value $c_1^\ast$, found by
setting $\dot c_1 = 0$ in Eq.~\eqref{eq:dc1}; for our parameters $c_1^\ast$
rises from $\approx 0.008$ at $p_1 = 0.1$ to $\approx 0.055$ at $p_1 = 0.4$.
Where $c_1^\ast$ is small, its repression of $c_2$ is negligible:
$R(c_1^\ast) \approx 1$, the neglected term $1 - R(c_1^\ast)$ growing from below
$10^{-3}$ at low $p_1$ to ${\sim}3\times10^{-2}$ by $p_1 = 0.4$.

We substitute $R(c_1^\ast) \to 1$ into Eq.~\eqref{eq:dc2} and set
$\dot c_2 = 0$. This gives a closed quadratic for the steady-state dominant
permeability,
\begin{equation}
  \frac{S_2}{c_{\max}}\,c_2^{\,2}
  - \bigl(S_2 - p_1\kappa\bigr)\,c_2
  - p_1\kappa\, b = 0,
  \label{eq:quadratic}
\end{equation}
whose physical root $c_2^\ast(p_1)$ is the one-trait prediction.
The one- and two-trait predictions agree to within $2\times10^{-3}$
(Table~S4 of the Supplementary Material). The discrepancy grows as $p_1 \to 0.5$, until at
the symmetric point the assumption of a negligible subordinate trait no longer
holds and the reduction ceases to apply.
%

The quadratic Eq.~\eqref{eq:quadratic} yields four results in closed form, each
with a direct counterpart in the IBM (Sec.~\ref{sec:results_threshold}).

First, in the deterministic, mutation-free limit the disfavored trait is lost
below a critical frequency $p_1^{\rm crit} \approx 0.17$, where the relaxation
pressure $(1-p_1)\kappa$ overtakes the selection gradient. This is a
transcritical bifurcation; retaining the full two-trait coupling places it at
$p_1 \approx 0.24$. At finite mutation, the drift floor $b$ unfolds the
bifurcation: the subordinate trait is never exactly lost but held near $b$,
rising smoothly through the crossover seen in the IBM near $p_1 \approx 0.25$
(Fig.~\ref{fig:steady_state_phi}A), in good agreement with the two-trait value.

Second, the symmetry relation $c_2^\ast(p_1) = c_1^\ast(1-p_1)$ holds because
each trait's steady state depends on $p_1$ only through $p_i$, with
$p_j = 1-p_i$. The IBM confirms it: $c_2 = 0.101$ at $p_1 = 0.3$ versus
$c_1 = 0.100$ at $p_1 = 0.7$.

Third, the two branches meet at $p_1 = 0.5$ at the coexistence value
$c^\ast = 0.072$.

Fourth, total permeability is nearly conserved. Setting $b\to 0$ and
$R\approx 1$ in Eq.~\eqref{eq:quadratic} gives
$c_i^{*} \approx c_{\max}(1 - p_j\kappa/S_i)$. For $p_i \gg K$ the Hill term
saturates, $S_i \approx A \equiv \lambda_0 f_{\rm cross}(\sigma_{\rm ON})$,
and since $p_1+p_2=1$,
\begin{equation}
c_1^{*}+c_2^{*} \approx c_{\max}\left(2-\frac{\kappa}{A}\right),
\label{eq:budget}
\end{equation}
independent of $p_1$. Each trait falls short of $c_{\max}$ in proportion to
the frequency of the other environment. The two frequencies sum to one, so
the shortfalls sum to a constant. Eq.~\eqref{eq:budget} gives $0.145$ at
$\sigma_{\rm ON}=50$. The full model gives $0.135$--$0.144$ and the IBM
$0.139$--$0.142$ over $p_1=0.1$--$0.5$. The approximation fails below
$p_1\approx 0.2$, where $S_1$ is not saturated and the subordinate trait
falls to the drift floor. Eq.~\eqref{eq:budget} also predicts that the
budget peaks at $\sigma^{*}$. The IBM sum at $p_1=0.2$ follows this:
$0.125$, $0.139$, $0.142$ and $0.135$ at $\sigma_{\rm ON}=35$, 50, 66 and 100.

The reduction reproduces the dominant permeability across the full range of
$p_1$ (Fig.~\ref{fig:unified_2d}A). It is most accurate near $p_1 \approx 0.3$
and deviates by up to $5\%$ at the extremes. There the disfavored trait is
small but non-zero, and the spatial collective field, which the mean-field
reduction omits, likely contributes to the deviation.
\subsection{Coexistence at the symmetric point}
\label{sec:model_bistable}
At $p_1 = 0.5$ the one-trait reduction no longer applies. It assumes a
subordinate trait pinned near the drift floor, whose repression of the
dominant trait is negligible. At symmetry both traits rise to comparable
levels and repress each other appreciably, so the full system
\eqref{eq:dc1}--\eqref{eq:dc2} must be retained. Under symmetric noise the
dynamics are invariant under $c_1 \leftrightarrow c_2$. With the weak cross-repression adopted for the model (Sec.~XI of the Supplementary Material), the system has a single stable fixed point, on the $c_1 = c_2$ diagonal at $(0.072, 0.072)$ (Fig.~\ref{fig:unified_2d}C): both permeabilities are
maintained at equal, intermediate levels.

Which structure obtains depends on the strength of cross-repression. Strong repression would split the diagonal point into two off-diagonal fixed points, giving winner-take-all bistability; weak repression yields the single coexistence point. 
The empirical gradient does not distinguish the two cases
(Sec.~XI of the Supplementary Material), so the choice is settled by the IBM outcome:
replicate trials converge on the diagonal rather than to either axis.
In the IBM, independent trials at $p_1 = 0.5$
converge toward a state with both traits elevated and comparable, near the predicted location (Sec.~\ref{sec:results_bistability}). The approach is
slow and stochastic, only ${\sim}40\%$ of trials converge by generation $10{,}000$, consistent with the weak deterministic gradient near the fixed
point.

When the two channels carry different sensing noise ($\sigma_{\rm ON}^{(1)} \neq \sigma_{\rm ON}^{(2)}$), the per-trait factor
$f_{\rm cross}(\sigma_{\rm ON})$ in Eq.~\eqref{eq:selection} breaks the $c_1 \leftrightarrow c_2$ symmetry. The fixed point remains single and stable but moves off the diagonal, toward the channel whose noise lies closer to the saddle-crossing optimum $\sigma^* \approx 66$. For $\sigma_{\rm ON}^{(1)} = 35$, $\sigma_{\rm ON}^{(2)} = 66$, the larger crossing fraction of
the $c_2$ channel ($f_{\rm cross}(66) = 0.209$ versus $f_{\rm cross}(35) = 0.149$) places the fixed point at $(c_1, c_2) = (0.051, 0.078)$. The IBM gives $c_1 \approx 0.060$, $c_2 \approx 0.074$ (Sec.~\ref{sec:results_bistability}), the same
direction of bias, weaker in magnitude. Noise asymmetry raises the optimal-noise channel and the fixed point does not split.
\section{Discussion}
\label{sec:discussion}

Two factors shape the evolution of quorum sensing in our model. One is how often each environment occurs. The other is how much noise each cell has in
sensing it. Environmental frequency decides how a nearly fixed investment is divided between two signals. Sensing noise decides how large that investment is,
through an interior optimum. At equal frequencies, both signals persist.

A single cell cannot reliably tell whether the current environment warrants a costly phenotype switch. In our model, quorum sensing solves this through
a collective decision rather than better individual sensing. The shared autoinducer field returns cells that have committed to the wrong attractor
basin to the majority state. Because each cell starts near the correct state, this is error correction rather than pooling of independent evidence
(Sec.~\ref{sec:limitations}). It evolves at every environmental frequency, while co-adaptation of both traits is confined to intermediate frequencies.

The main difference from the single-autoinducer model of Ref.~\citep{MorenoGamez2023} is the toggle. Cells must cross a saddle-point
separatrix to switch states, so they can become stuck in the wrong basin. Once the population crosses together, positive feedback makes the new state
self-reinforcing. The evolved permeability balances the benefit of this collective feedback against its metabolic cost and the relaxation of the
inactive locus.

The subordinate trait is never lost, but it is strongly suppressed when its environment is rare. At $p_1 = 0.2$, $c_2$ reaches $0.119$ while $c_1$
settles at $0.020$. Two asymmetries act together. The subordinate locus relaxes four times faster ($p_2\rho = 0.0016$ against $p_1\rho = 0.0004$
per generation). It also pays the metabolic cost in every generation but is rewarded only when its own environment occurs. At $p_1 = 1$ relaxation vanishes, yet the trait still plateaus at $c_{\max} = 0.140$, where the tracking gain no longer repays the cost.
Suppression weakens toward symmetry, and by $p_1 = 0.3$ the subordinate clears the drift boundary. In the deterministic limit, the balance between
selection and relaxation gives a transcritical bifurcation near $p_1 \approx 0.24$. Mutation unfolds it into the smooth crossover seen in the IBM.

The two traits also share an approximately fixed budget. Across $p_1 = 0.1$--$0.5$, $c_1 + c_2$ stays between $0.139$ and $0.142$, within
$1.5\%$ of $c_{\max}$, although environmental entropy more than doubles over the same range. Environmental frequency sets how the budget is split, not
its size. Each trait falls short of $c_{\max}$ in proportion to the frequency of the other environment, and the two frequencies sum to one
(Eq.~\ref{eq:budget}).
These results bear on why bacteria carry several QS systems. \textit{Vibrio harveyi} runs three parallel autoinducer channels that
converge on a shared output~\cite{Henke2004, Long2009}. Distinct signals can report distinct features of the environment~\cite{Cornforth2014}, and
redundant systems can accumulate through social cheating~\cite{EvenTov2016}. Our model gives a third route that needs neither. A channel tied to a rare
environment is kept at a low but non-zero level by fluctuating selection. Across strains or habitats, the relative investment in each channel should
then track how often its environment occurs, while total investment varies much less.

The same balance gives a negative case. QS is prominent in bacteria whose environments change repeatedly, such as pathogens cycling through
infection~\citep{Bassler2002}. Where an environment is rarely encountered, its channel decays to the mutation--drift floor, even if initially present.
A QS locus gained by horizontal transfer should therefore persist only in recipients that meet the matching environment often enough. Comparing
transfer events with the environmental variability of recipient lineages could test whether this contributes to the patchy phylogenetic distribution
of QS~\citep{Crespi2001}.

Suppression is also reversible, and loss is faster than gain. After a shift from $p_1 = 0.2$ to $0.8$, $c_2$ would decay by relaxation within
$1/(p_1\rho) \approx 600$ generations, while $c_1$ would take several relaxation times to rise (Fig.~\ref{fig:steady_state_phi}B). A population
tracking a slowly drifting environment should therefore lag more when an environment becomes common than when it becomes rare. Experimental evolution
under ramped environmental frequencies could test this.

Sensing noise shapes the evolved permeability through an interior optimum (Fig.~\ref{fig:unified_2d}B). Sensing error places cells below the toggle
saddle, and their rescue by the shared field is what makes permeability worth paying for. With too little error there is little to correct. With
too much, truncation at zero pushes the extra spread above the active state, and fewer cells fall below the saddle. The toggle geometry places
the optimum at $\sigma^{*} \approx 66$ (Sec.~\ref{sec:model_noise}), consistent with the IBM.

This optimum differs from earlier results on collective sensing. In Ref.~\cite{MorenoGamez2023}, more sensing noise favors collective sensing
provided cells are correct on average. Salahshour found an optimum in communication noise near a bistable edge, with all noise levels fixed~\cite{Salahshour2019PRL}. Here the optimum lies in sensing noise and appears in an evolved trait. Cells stay correct on average at every noise
level, so the reversal comes from truncation, not from biased estimates.

The optimum also parallels stochastic resonance, where intermediate noise maximizes signal detection in a threshold system~\citep{Gammaitoni1998}.
The toggle saddle plays the threshold and sensing error the noise, although the environmental switching here is not periodic. The key difference is who
benefits. A cell that mis-commits gains nothing from its own error. The gain goes to the population, through the shared field that returns mis-committed
cells to the majority state. The optimum is therefore a property of the population, not of any single sensor.

At equal environmental frequencies both traits evolve and coexist. What varies between replicate populations is the timing of adaptation, not the
identity of a winner. One trait usually crosses the drift boundary first, and the other follows. Noise asymmetry does not turn this into fixation. It
grades the level at which the two traits coexist, with the channel nearer $\sigma^{*}$ settling higher. The two-trait model has a single stable node
on the diagonal (Fig.~\ref{fig:unified_2d}C). Weak cross-repression keeps the two basins merged, and stronger repression would split them into
competing attractors. The repression strength is not measured, so the IBM, not the model, settles which structure applies.

Maintaining both channels resembles bet hedging, and its cost is the cost of the information the population buys. Theory on fluctuating environments
quantifies this trade-off~\citep{Kussell2005, DonaldsonMatasci2010, Rivoire2011}, and phenotypic bet hedging is widespread among
bacteria~\citep{Morawska2022}. Our result adds that the hedge need not be stochastic switching between phenotypes. It can be carried by two signaling
channels, with selection leaving the split between them soft.

One adaptive-dynamics model covers both regimes. Away from symmetry the disfavored trait decouples, and the model reduces to a closed one-dimensional equation for the dominant permeability. At $p_1 = 0.5$ both traits rise to comparable levels, and the full two-dimensional dynamics give
a single coexistence attractor. The model reproduces both traits across all $p_1$ (Fig.~\ref{fig:unified_2d}A). Its only finite-population input is the
drift floor $b$, and this sets the subordinate level. Without it the disfavored trait would sit at zero. With it, the model matches the IBM
subordinate, for example $c_1 \approx 0.020$ at $p_1 = 0.2$ in both. Mutation therefore sets the baseline from which the disfavored trait is
lifted, rather than adding a residual the deterministic theory misses.
The model also predicts that the noise optimum does not depend on environmental frequency. Sensing noise enters only through $f_{\rm cross}$,
so $\sigma^{*} \approx 66$ for every $p_1$ and both traits. The size of the effect should grow toward symmetry. In the model, the dominant trait gains
$9\%$ between $\sigma_{\rm ON} = 35$ and $66$ at $p_1 = 0.2$, and $24\%$ at $p_1 = 0.4$.

\subsection{Limitations and future directions}
\label{sec:limitations}

The most important limitation concerns what cells sense. Each cell starts with its active channel centered on the correct target, so sensing noise is an error in commitment, not uncertainty about the environment. The shared field, therefore, corrects errors rather than pooling independent evidence. A model with ambiguous environmental cues would test collective inference properly.

Several results rest on fixed or assumed ingredients. The value $\sigma^{*} \approx 66$ follows from this toggle's geometry, which we did not vary; only the existence of the optimum is general. The relaxation rate
$\rho$ is imposed, not derived, and it sets the subordinate level, the loss timescale and the budget. Population size, mutation rate and cost were held fixed.

The adaptive-dynamics model omits the spatial field. It deviates from the IBM by up to ${\sim}5\%$ at the frequency extremes, misses the shape of the plateau above $\sigma^{*}$, and underestimates adaptation time near symmetry by about $2.5$-fold. Its cross-repression strength is adopted (Sec.~XI of the Supplementary Material), and it holds $\sigma_{\rm OFF}$ fixed, although the subordinate trait rises fifteenfold with it (Sec.~V of the
Supplementary Material). At $p_1 = 0.5$, about $60\%$ of trials remain near the drift floor after $10{,}000$ generations, so the coexistence level is read from resolved trials.

Natural extensions include networks of more than two autoinducers, as in \textit{Vibrio harveyi}; environments correlated in time, where populations may track rather than average~\cite{Kussell2005}; and cheaters that defect on one channel only, extending single-signal work~\cite{MorenoGamez2023}.
The framework makes two falsifiable predictions. Total investment in the two signals should stay nearly constant across environmental frequencies while their ratio shifts (Eq.~\ref{eq:budget}). The dominant permeability should fall linearly with the frequency of the competing environment. Evolve-and-resequence experiments under controlled switching would test both.
\section{Conclusions}
\label{sec:conclusions}

We studied how QS permeability evolves in fluctuating environments, using an individual-based model of a two-autoinducer toggle switch coupled to a shared extracellular field. Permeability evolves through a collective mechanism: high-permeability cells amplify the shared field and rescue neighbors from the wrong attractor basin. Three findings follow.

First, environmental frequency sets how a nearly fixed total permeability is divided between the two traits. The trait tied to a rare environment is suppressed toward the mutation--drift baseline but not lost. Its onset is a
smooth crossover, sharp only in the deterministic limit, where it is a transcritical bifurcation that mutation unfolds. Toward equal frequencies, the two plateaus converge while the time to reach them lengthens. Suppression
also requires low inactive-channel noise.

Second, sensing noise is required, and its effect is non-monotonic. Without noise no cell mis-commits and neither permeability evolves. With noise, the evolved plateau has an interior optimum in $\sigma_{\rm ON} \approx 50$--$66$,
consistent with the saddle-crossing optimum $\sigma^{*} \approx 66$ set by the toggle geometry. Adaptation speed instead rises with noise and then saturates. Noise therefore controls speed and plateau differently.

Third, equal environmental frequencies produce stochastic coexistence. Both traits evolve and persist; the stochasticity lies in the timing of adaptation, not in the identity of a winner. Noise asymmetry between the two
channels grades the coexistence level: the channel nearer $\sigma^{*}$ adapts faster and settles higher.

A two-trait adaptive-dynamics model with weak cross-repression connects these results. Away from symmetry it reduces to a closed one-dimensional equation that reproduces the IBM across the asymmetric range ($R^2 = 0.998$). It yields the crossover, the symmetry relation and the conserved budget in closed form. At $p_1 = 0.5$ the full two-dimensional dynamics have a single coexistence attractor, which is why replicate
populations keep both traits.

Together, these results show when a population can reach a clear collective decision. Under strong environmental bias, roughly three to one or greater, the population commits within a few relaxation times. One channel evolves to a high plateau and the other is held near the drift floor. For $0.3 \lesssim p_1 \lesssim 0.7$, the population instead maintains both channels at frequency-graded levels. This hedge is slow to establish, and slowest at equal frequencies.
\begin{acknowledgments}
We thank Valentin Anfray for insightful discussions and suggestions. This work was supported by grants from the National Science and Technology Council, Taiwan (Grant No. NSTC 111-2112-M-001-027-MY3, 114-2112-M-001-062 and 115-2112-M-001 -055) and Academia Sinica Career Development Award (Project No. AS-CDA-114-M02).
\end{acknowledgments}

\bibliographystyle{apsrev4-2}
\bibliography{qs}

@article{Flemming2016,
  author  = {Flemming, H.-C. and Wingender, J. and Szewzyk, U. and Steinberg, P. and Rice, S. A. and Kjelleberg, S.},
  title   = {Biofilms: an emergent form of bacterial life},
  journal = {Nature Reviews Microbiology},
  year    = {2016},
  volume  = {14},
  number  = {9},
  pages   = {563--575},
  doi     = {10.1038/nrmicro.2016.94}
}

@article{HallStoodley2004,
  author  = {Hall-Stoodley, L. and Costerton, J. W. and Stoodley, P.},
  title   = {Bacterial biofilms: from the natural environment to infectious diseases},
  journal = {Nature Reviews Microbiology},
  year    = {2004},
  volume  = {2},
  number  = {2},
  pages   = {95--108},
  doi     = {10.1038/nrmicro821}
}

@article{Balaban2004,
  author  = {Balaban, N. Q. and Merrin, J. and Chait, R. and Kowalik, L. and Leibler, S.},
  title   = {Bacterial persistence as a phenotypic switch},
  journal = {Science},
  year    = {2004},
  volume  = {305},
  number  = {5690},
  pages   = {1622--1625},
  doi     = {10.1126/science.1099390}
}

@article{Kussell2005,
  author  = {Kussell, E. and Leibler, S.},
  title   = {Phenotypic diversity, population growth, and information in fluctuating environments},
  journal = {Science},
  year    = {2005},
  volume  = {309},
  number  = {5743},
  pages   = {2075--2078},
  doi     = {10.1126/science.1114383}
}

@article{Henke2004,
  author  = {Henke, Jennifer M. and Bassler, Bonnie L.},
  title   = {Three parallel quorum-sensing systems regulate gene expression
             in \textit{Vibrio harveyi}},
  journal = {J. Bacteriol.},
  volume  = {186},
  number  = {20},
  pages   = {6902--6914},
  year    = {2004},
  doi     = {10.1128/JB.186.20.6902-6914.2004}
}

@article{Long2009,
  author  = {Long, Tao and Tu, Kimberly C. and Wang, Yufang and Mehta, Pankaj
             and Ong, N. P. and Bassler, Bonnie L. and Wingreen, Ned S.},
  title   = {Quantifying the integration of quorum-sensing signals with
             single-cell resolution},
  journal = {PLoS Biol.},
  volume  = {7},
  number  = {3},
  pages   = {e1000068},
  year    = {2009},
  doi     = {10.1371/journal.pbio.1000068}
}

@article{Quan2016,
  author  = {Quan, David N. and Tsao, Chen-Yu and Wu, Hsuan-Chen
             and Bentley, William E.},
  title   = {Quorum sensing desynchronization leads to bimodality and
             patterned behaviors},
  journal = {PLoS Comput. Biol.},
  volume  = {12},
  number  = {4},
  pages   = {e1004781},
  year    = {2016},
  doi     = {10.1371/journal.pcbi.1004781}
}

@article{EvenTov2016,
  author  = {Even-Tov, Eran and Omer Bendori, Shira and Valastyan, Julie
             and Ke, Xiaobo and Pollak, Shaul and Bareia, Tasneem
             and Ben-Zion, Ishay and Bassler, Bonnie L. and Eldar, Avigdor},
  title   = {Social evolution selects for redundancy in bacterial quorum
             sensing},
  journal = {PLoS Biol.},
  volume  = {14},
  number  = {2},
  pages   = {e1002386},
  year    = {2016},
  doi     = {10.1371/journal.pbio.1002386}
}

@article{Beaumont2009,
  author  = {Beaumont, H. J. E. and Gallie, J. and Kost, C. and Ferguson, G. C. and Rainey, P. B.},
  title   = {Experimental evolution of bet hedging},
  journal = {Nature},
  year    = {2009},
  volume  = {462},
  number  = {7269},
  pages   = {90--93},
  doi     = {10.1038/nature08504}
}

@article{Rutherford2012,
  author  = {Rutherford, S. T. and Bassler, B. L.},
  title   = {Bacterial quorum sensing: its role in virulence and possibilities for its control},
  journal = {Cold Spring Harbor Perspectives in Medicine},
  year    = {2012},
  volume  = {2},
  number  = {11},
  pages   = {a012427},
  doi     = {10.1101/cshperspect.a012427}
}

@article{Whiteley2017,
  author  = {Whiteley, M. and Diggle, S. P. and Greenberg, E. P.},
  title   = {Progress in and promise of bacterial quorum sensing research},
  journal = {Nature},
  year    = {2017},
  volume  = {551},
  number  = {7680},
  pages   = {313--320},
  doi     = {10.1038/nature24624}
}

@article{Mukherjee2019,
  author  = {Mukherjee, S. and Bassler, B. L.},
  title   = {Bacterial quorum sensing in complex and dynamically changing environments},
  journal = {Nature Reviews Microbiology},
  year    = {2019},
  volume  = {17},
  number  = {6},
  pages   = {371--382},
  doi     = {10.1038/s41579-019-0186-5}
}

@article{Keller2006,
  author  = {Keller, L. and Surette, M. G.},
  title   = {Communication in bacteria: an ecological and evolutionary perspective},
  journal = {Nature Reviews Microbiology},
  year    = {2006},
  volume  = {4},
  number  = {4},
  pages   = {249--258},
  doi     = {10.1038/nrmicro1383}
}

@article{Bassler2002,
  author  = {Bassler, Bonnie L.},
  title   = {Small Talk: Cell-to-Cell Communication in Bacteria},
  journal = {Cell},
  volume  = {109},
  number  = {4},
  pages   = {421--424},
  year    = {2002},
  doi     = {10.1016/S0092-8674(02)00749-3},
}

@article{Crespi2001,
  author  = {Crespi, Bernard J.},
  title   = {The evolution of social behavior in microorganisms},
  journal = {Trends in Ecology \& Evolution},
  year    = {2001},
  volume  = {16},
  number  = {4},
  pages   = {178--183},
  doi     = {10.1016/S0169-5347(01)02113-2}
}

@article{Ng2009,
  author  = {Ng, W.-L. and Bassler, B. L.},
  title   = {Bacterial quorum-sensing network architectures},
  journal = {Annual Review of Genetics},
  year    = {2009},
  volume  = {43},
  pages   = {197--222},
  doi     = {10.1146/annurev-genet-102108-134304}
}

@article{Gammaitoni1998,
  author  = {Gammaitoni, Luca and H{\"a}nggi, Peter
             and Jung, Peter and Marchesoni, Fabio},
  title   = {Stochastic resonance},
  journal = {Reviews of Modern Physics},
  year    = {1998},
  volume  = {70},
  number  = {1},
  pages   = {223--287},
  doi     = {10.1103/RevModPhys.70.223}
}

@article{Cao2017Ready,
   author       = {Cao, Mengyi and Goodrich-Blair, Heidi},
   title        = {Ready or Not: Microbial Adaptive Responses in Dynamic Symbiosis Environments},
   journal      = {Journal of Bacteriology},
   year         = {2017},
   volume       = {199},
   number       = {15},
   pages        = {e00883-16},
   doi          = {10.1128/JB.00883-16},
}

@article{West2012,
   author = {Stuart A. West and Klaus Winzer and Andy Gardner and Stephen P. Diggle},
   doi = {10.1016/j.tim.2012.09.004},
   issn = {0966842X},
   issue = {12},
   journal = {Trends in Microbiology},
   month = {12},
   pages = {586-594},
   pmid = {23084573},
   title = {Quorum sensing and the confusion about diffusion},
   volume = {20},
   url = {https://linkinghub.elsevier.com/retrieve/pii/S0966842X12001746},
   year = {2012}
}

@article{Wu2014quorum,
  title = {Quorum-sensing crosstalk-driven synthetic circuits: from unimodality to trimodality},
  author = {Wu, Fuqing and Menn, David J. and Wang, Xiao},
  journal = {Chemistry \& Biology},
  volume = {21},
  number = {12},
  pages = {1629--1638},
  year = {2014},
  month = dec,
  doi = {10.1016/j.chembiol.2014.10.008},
}

@article{Nikolaev2016,
   author = {Evgeni V. Nikolaev and Eduardo D. Sontag},
   doi = {10.1371/journal.pcbi.1004881},
   issn = {15537358},
   issue = {4},
   journal = {PLoS Computational Biology},
   month = {4},
   pages = {e1004881},
   pmid = {27128344},
   publisher = {Public Library of Science},
   title = {Quorum-Sensing Synchronization of Synthetic Toggle Switches: A Design Based on Monotone Dynamical Systems Theory},
   volume = {12},
   year = {2016}
}

@article{Weber2016,
   author = {Marc Weber and Javier Buceta},
   doi = {10.1098/rsif.2015.1092},
   issn = {1742-5689},
   issue = {119},
   journal = {Journal of The Royal Society Interface},
   month = {6},
   pages = {20151092},
   pmid = {27307510},
   publisher = {Royal Society of London},
   title = {The cellular {Ising model}: a framework for phase transitions in multicellular environments},
   volume = {13},
   url = {https://royalsocietypublishing.org/doi/10.1098/rsif.2015.1092},
   year = {2016}
}

@article{MorenoGamez2023,
   author = {Stefany Moreno-G{\'a}mez and Michael E. Hochberg and G. S. van Doorn},
   doi = {10.1038/s41467-023-37950-7},
   issn = {2041-1723},
   number = {1},
   journal = {Nature Communications},
   month = {jun},
   pages = {3415},
   pmid = {37296108},
   publisher = {Nature Research},
   title = {Quorum sensing as a mechanism to harness the wisdom of the crowds},
   volume = {14},
   url = {https://www.nature.com/articles/s41467-023-37950-7},
   year = {2023},
}

@article{Miller2001,
   author = {Miller, Melissa B. and Bassler, Bonnie L.},
   doi = {10.1146/annurev.micro.55.1.165},
   issn = {0066-4227},
   issue = {1},
   journal = {Annual Review of Microbiology},
   month = {October},
   pages = {165-199},
   title = {Quorum Sensing in Bacteria},
   volume = {55},
   url = {https://www.annualreviews.org/doi/10.1146/annurev.micro.55.1.165},
   year = {2001}
}

@article{Papenfort2016,
   author = {Papenfort, Kai and Bassler, Bonnie L.},
   doi = {10.1038/nrmicro.2016.89},
   issn = {1740-1526},
   issue = {9},
   journal = {Nature Reviews Microbiology},
   month = {September},
   pages = {576-588},
   pmid = {27510864},
   publisher = {Nature Publishing Group},
   title = {Quorum sensing signal-response systems in Gram-negative bacteria},
   volume = {14},
   url = {https://www.nature.com/articles/nrmicro.2016.89},
   year = {2016}
}

@article{Darch2012,
   author = {Sophie E. Darch and Stuart A. West and Klaus Winzer and Stephen P. Diggle},
   doi = {10.1073/pnas.1118131109},
   issn = {0027-8424},
   issue = {21},
   journal = {Proceedings of the National Academy of Sciences},
   month = {5},
   pages = {8259-8263},
   pmid = {22566647},
   title = {Density-dependent fitness benefits in quorum-sensing bacterial populations},
   volume = {109},
   url = {https://pnas.org/doi/full/10.1073/pnas.1118131109},
   year = {2012}
}

@article{Schuster2017,
   author = {Martin Schuster and D. Joseph Sexton and Burkhard A. Hense},
   doi = {10.3389/fmicb.2017.00885},
   issn = {1664-302X},
   journal = {Frontiers in Microbiology},
   month = {5},
   pages = {885},
   publisher = {Frontiers Media S.A.},
   title = {Why Quorum Sensing Controls Private Goods},
   volume = {8},
   url = {http://journal.frontiersin.org/article/10.3389/fmicb.2017.00885/full},
   year = {2017}
}

@article{Decho2010,
   author = {Alan W. Decho and R. Sean Norman and Pieter T. Visscher},
   doi = {10.1016/j.tim.2009.12.008},
   issn = {0966842X},
   issue = {2},
   journal = {Trends in Microbiology},
   month = {2},
   pages = {73-80},
   pmid = {20060299},
   title = {Quorum sensing in natural environments: emerging views from microbial mats},
   volume = {18},
   url = {https://linkinghub.elsevier.com/retrieve/pii/S0966842X09002674},
   year = {2010}
}

@article{Waters2005,
   author = {Christopher M. Waters and Bonnie L. Bassler},
   doi = {10.1146/annurev.cellbio.21.012704.131001},
   issn = {1081-0706},
   issue = {1},
   journal = {Annual Review of Cell and Developmental Biology},
   month = {11},
   pages = {319-346},
   pmid = {16212498},
   title = {QUORUM SENSING: Cell-to-Cell Communication in Bacteria},
   volume = {21},
   url = {https://www.annualreviews.org/doi/10.1146/annurev.cellbio.21.012704.131001},
   year = {2005},
}

@article{Jemielita2018,
   author = {Matthew Jemielita and Ned S Wingreen and Bonnie L Bassler},
   doi = {10.7554/eLife.42057},
   issn = {2050-084X},
   journal = {eLife},
   month = {12},
   pages = {e42057},
   title = {Quorum sensing controls {\textit{Vibrio cholerae}} multicellular aggregate formation},
   volume = {7},
   url = {https://elifesciences.org/articles/42057},
   year = {2018},
}

@article{Morawska2022,
   author = {Luiza P. Morawska and Jhonatan A. Hernandez-Valdes and Oscar P. Kuipers},
   doi = {10.1002/wsbm.1544},
   issn = {2692-9368},
   issue = {2},
   journal = {WIREs Mechanisms of Disease},
   month = {3},
   pages = {e1544},
   pmid = {35266649},
   publisher = {John Wiley and Sons Inc},
   title = {Diversity of bet-hedging strategies in microbial communities---Recent cases and insights},
   volume = {14},
   url = {https://wires.onlinelibrary.wiley.com/doi/10.1002/wsbm.1544},
   year = {2022},
}

@article{Bareia2017,
   author = {Tasneem Bareia and Shaul Pollak and Avigdor Eldar},
   doi = {10.1038/s41564-017-0044-z},
   issn = {2058-5276},
   issue = {1},
   journal = {Nature Microbiology},
   month = {10},
   pages = {83-89},
   pmid = {29038467},
   publisher = {Nature Publishing Group},
   title = {Self-sensing in {\textit{Bacillus subtilis}} quorum-sensing systems},
   volume = {3},
   url = {https://www.nature.com/articles/s41564-017-0044-z},
   year = {2017},
}

@article{Schultz2009,
  author  = {Schultz, D. and Wolynes, P. G. and Ben-Jacob, E. and Onuchic, J. N.},
  title   = {Deciding fate in adverse times: sporulation and competence in {Bacillus subtilis}},
  journal = {Proceedings of the National Academy of Sciences of the United States of America},
  year    = {2009},
  volume  = {106},
  number  = {50},
  pages   = {21027--21034},
  doi     = {10.1073/pnas.0912185106}
}

@article{Maamar2005,
  author  = {Maamar, H. and Dubnau, D.},
  title   = {Bistability in the {Bacillus subtilis} {K}-state (competence) system requires a positive feedback loop},
  journal = {Molecular Microbiology},
  year    = {2005},
  volume  = {56},
  number  = {3},
  pages   = {615--624},
  doi     = {10.1111/j.1365-2958.2005.04592.x}
}

@article{Veening2005,
  author  = {Veening, J.-W. and Hamoen, L. W. and Kuipers, O. P.},
  title   = {Phosphatases modulate the bistable sporulation gene expression pattern in {Bacillus subtilis}},
  journal = {Molecular Microbiology},
  year    = {2005},
  volume  = {56},
  number  = {6},
  pages   = {1481--1494},
  doi     = {10.1111/j.1365-2958.2005.04659.x}
}

@article{Mellbye2014,
  author  = {Mellbye, B. and Schuster, M.},
  title   = {Physiological framework for the regulation of quorum sensing-dependent public goods in {Pseudomonas aeruginosa}},
  journal = {Journal of Bacteriology},
  year    = {2014},
  volume  = {196},
  number  = {6},
  pages   = {1155--1164},
  doi     = {10.1128/JB.01223-13}
}

@article{Leung2012,
  author  = {Leung, V. and L{\'e}vesque, C. M.},
  title   = {A stress-inducible quorum-sensing peptide mediates the formation of persister cells with noninherited multidrug tolerance},
  journal = {Journal of Bacteriology},
  year    = {2012},
  volume  = {194},
  number  = {9},
  pages   = {2265--2274},
  doi     = {10.1128/JB.06707-11}
}

@article{An2014,
  author  = {An, J. H. and Goo, E. and Kim, H. and Seo, Y.-S. and Hwang, I.},
  title   = {Bacterial quorum sensing and metabolic slowing in a cooperative population},
  journal = {Proceedings of the National Academy of Sciences of the United States of America},
  year    = {2014},
  volume  = {111},
  number  = {41},
  pages   = {14912--14917},
  doi     = {10.1073/pnas.1412431111}
}

@article{Cornforth2014,
  author  = {Cornforth, D. M. and Popat, R. and McNally, L. and Gurney, J.
             and Scott-Phillips, T. C. and Ivens, A. and Diggle, S. P.
             and Brown, S. P.},
  title   = {Combinatorial quorum sensing allows bacteria to resolve their social and physical environment},
  journal = {Proceedings of the National Academy of Sciences of the United States of America},
  year    = {2014},
  volume  = {111},
  number  = {11},
  pages   = {4280--4284},
  doi     = {10.1073/pnas.1319175111}
}

@article{Danino2010,
  author  = {Danino, Tal and Mondrag{\'o}n-Palomino, Octavio and Tsimring, Lev and Hasty, Jeff},
  title   = {A synchronized quorum of genetic clocks},
  journal = {Nature},
  volume  = {463},
  number  = {7279},
  pages   = {326--330},
  year    = {2010},
  doi     = {10.1038/nature08753}
}

@article{Gardner2000,
  author  = {Gardner, Timothy S. and Cantor, Charles R. and Collins, James J.},
  title   = {Construction of a genetic toggle switch in {\textit{Escherichia coli}}},
  journal = {Nature},
  volume  = {403},
  number  = {6767},
  pages   = {339--342},
  year    = {2000},
  doi     = {10.1038/35002131}
}

@article{Redfield2002,
  author  = {Redfield, R. J.},
  title   = {Is quorum sensing a side effect of diffusion sensing?},
  journal = {Trends in Microbiology},
  year    = {2002},
  volume  = {10},
  number  = {8},
  pages   = {365--370},
  doi     = {10.1016/S0966-842X(02)02400-9}
}

@article{Berdahl2013,
  author  = {Berdahl, A. and Torney, C. J. and Ioannou, C. C. and Faria, J. J. and Couzin, I. D.},
  title   = {Emergent sensing of complex environments by mobile animal groups},
  journal = {Science},
  year    = {2013},
  volume  = {339},
  number  = {6119},
  pages   = {574--576},
  doi     = {10.1126/science.1225883}
}

@article{Golub2010,
  author  = {Golub, B. and Jackson, M. O.},
  title   = {Na{\"i}ve learning in social networks and the wisdom of crowds},
  journal = {American Economic Journal: Microeconomics},
  year    = {2010},
  volume  = {2},
  number  = {1},
  pages   = {112--149},
  doi     = {10.1257/mic.2.1.112}
}

@article{Kobayashi2004,
  author  = {Kobayashi, H. and K{\ae}rn, M. and Araki, M. and Chung, K.
             and Gardner, T. S. and Cantor, C. R. and Collins, J. J.},
  title   = {Programmable cells: interfacing natural and engineered gene networks},
  journal = {Proceedings of the National Academy of Sciences of the United States of America},
  year    = {2004},
  volume  = {101},
  number  = {22},
  pages   = {8414--8419},
  doi     = {10.1073/pnas.0402940101}
}

@article{Salahshour2019SciRep,
  author  = {Salahshour, Mohammad and Rouhani, Shahin and Roudi, Yasser},
  title   = {Phase transitions and asymmetry between signal comprehension
             and production in biological communication},
  journal = {Sci. Rep.},
  volume  = {9},
  number  = {1},
  pages   = {3428},
  year    = {2019},
  doi     = {10.1038/s41598-019-40141-4}
}

@article{Salahshour2019PRL,
  author  = {Salahshour, Mohammad},
  title   = {Phase Diagram and Optimal Information Use in a Collective
             Sensing System},
  journal = {Phys. Rev. Lett.},
  volume  = {123},
  number  = {6},
  pages   = {068101},
  year    = {2019},
  doi     = {10.1103/PhysRevLett.123.068101}
}

@article{Hense2007,
  author  = {Hense, B. A. and Kuttler, C. and M{\"u}ller, J. and Rothballer, M.
             and Hartmann, A. and Kreft, J.-U.},
  title   = {Does efficiency sensing unify diffusion and quorum sensing?},
  journal = {Nature Reviews Microbiology},
  year    = {2007},
  volume  = {5},
  number  = {3},
  pages   = {230--239},
  doi     = {10.1038/nrmicro1600}
}

@article{Rivoire2011,
  author  = {Rivoire, O. and Leibler, S.},
  title   = {The value of information for populations in varying environments},
  journal = {Journal of Statistical Physics},
  year    = {2011},
  volume  = {142},
  number  = {6},
  pages   = {1124--1166},
  doi     = {10.1007/s10955-011-0166-2}
}

@article{DonaldsonMatasci2010,
  author  = {Donaldson-Matasci, M. C. and Bergstrom, C. T. and Lachmann, M.},
  title   = {The fitness value of information},
  journal = {Oikos},
  year    = {2010},
  volume  = {119},
  number  = {2},
  pages   = {219--230},
  doi     = {10.1111/j.1600-0706.2009.17781.x}
}
\clearpage        


\end{document}